\documentclass{article}

\usepackage{arxiv}

\usepackage{graphicx}
\usepackage{dcolumn}
\usepackage{listings}

\usepackage{bm}
\usepackage[mathlines]{lineno}

\usepackage{makecell}
\usepackage{multirow}
\usepackage{hhline}
\usepackage[utf8]{inputenc}
\usepackage[T1]{fontenc}
\usepackage{mathptmx}
\usepackage{etoolbox}
\usepackage{xcolor}
\usepackage{color}
\usepackage{float}
\usepackage{hyperref}
\newcommand{\be}{\begin{equation}}
\newcommand{\ee}{\end{equation}}

\usepackage{amsmath}
\usepackage[numbers, sort&compress]{natbib}

\usepackage[utf8]{inputenc} 
\usepackage[T1]{fontenc}    
\usepackage{hyperref}       
\usepackage{url}            
\usepackage{booktabs}       
\usepackage{amsfonts}       
\usepackage{nicefrac}       
\usepackage{microtype}      
\usepackage{lipsum}
\usepackage{graphicx}
\graphicspath{ {./images/} }
\usepackage{hyperref} 
\hypersetup{colorlinks = true, 
	        linkcolor = blue, 
	        urlcolor = blue,
            citecolor = blue} 
\usepackage[misc]{ifsym}

\title{A novel spatial distribution method for wind farm parameterizations based on the Gaussian function}

\author{
 Bowen Du \\
  North China Electric Power University, \\Beijing 102206, PR China \\
   \And
   Qi Li \\
   Peking University, \\Beijing  100871, PR China\\
   \And
  Mingwei Ge $^{\textrm{\Letter}}$ \\
  North China Electric Power University, \\Beijing 102206, PR China \\
  \texttt{${\textrm{\Letter}}$ gemingwei@ncepu.edu.cn}\\
  \And
 Xintao Li \\
  North China Electric Power University, \\Beijing 102206, PR China \\
  \And
 Yongqian Liu \\
  North China Electric Power University, \\Beijing 102206, PR China \\
}

\begin{document}
\maketitle
\begin{abstract}
Wind farm parameterizations are crucial for quantifying the wind-farm atmosphere interaction, where wind turbines are typically modeled as elevated momentum sinks and sources of turbulent kinetic energy (TKE). These quantities must be properly distributed to the mesoscale grid and integrated into the governing equations. Existing parameterizations use a single-column method solely based on the relationship between the coordinates of the rotor center and the mesoscale grid. However, this method fails to account for the effects of different wind turbine positions within the grid, particularly neglecting the contributions of other turbines located at grid boundaries. This can easily lead to errors of the spatial distribution of the sink and source, thereby impacting the accuracy of mesoscale flow simulations. To this end, we propose a multi-column spatial distribution method based on the Gaussian function. This method distributes the sink and source to multiple vertical grid columns based on the grid weights, which are analytically determined by integrating the two-dimensional Gaussian function over the mesoscale grid. We have applied this approach to the classic Fitch model, proposed the improved Fitch-Gaussian model, and integrated it into the mesoscale Weather Research and Forecasting model. Using high-fidelity large-eddy simulation as a benchmark, we set up stand-alone wind turbine and wind farm cases with turbines located at grid boundaries and compared the performance of the proposed method with the single-column method. The results show that the proposed method captures the spatial distribution of the sink and source more accurately, with a higher correlation coefficient and lower normalized root mean square error. However, due to the inherent limitations of the Fitch model, the Fitch-Gaussian model faces challenges in accurately predicting the magnitude and streamwise evolution of velocity deficit and added TKE in wind-farm wakes. Nevertheless, the Fitch-Gaussian model better captures the overall spatial distribution patterns of velocity deficit and added TKE. Therefore, the proposed Gaussian-based multi-column spatial distribution method is recommended for future mesoscale wind farm simulations, especially when the influence of the wind turbine rotor spans multiple mesoscale grid columns.
\end{abstract}

\section{\label{sec1}Introduction}
Global wind power development is showing a trend towards clustered and large-scale base development \cite{1veers2019grand}. China, the European Union, the United States and other countries have built and planned large-scale wind power bases. In wind power bases, wind farms are concentrated and organized in clusters. Downstream wind farms operate within the wakes of upstream ones, leading to a significant reduction in power generation \cite{2lundquist2019costs}. Wind-farm wakes extend over large spatial scales, and their evolution is affected by numerous factors such as Coriolis force and atmospheric thermal stratification, involving multi-physical phenomena such as the internal boundary layer and gravity waves. Numerical simulation is currently the most important research tool for our understanding of wind-farm atmosphere interactions \cite{3ouro2025numerical}. Considering the computational cost and the range of spatial scales that can be simulated, the mesoscale model coupled with wind farm parameterizations has received extensive attention and plays an important role in simulating the interaction between wind farm clusters and the atmospheric boundary layer \cite{4volker2017prospects,5pryor2021wind}, as well as in quantifying the local climate impact of wind power bases \cite{6akhtar2021accelerating,7akhtar2022impacts}.

The wind farm parameterization (WFP) is an indispensable physical scheme for characterizing wind farm effects in mesoscale models. Improving its accuracy is of great significance for better characterizing wind-farm atmosphere interactions \cite{8porte2020wind}. WFP models generally view wind farms as elevated momentum sinks and sources of turbulent kinetic energy (TKE) \cite{9fischereit2022review}. Their integration with mesoscale models consists of two main steps: the first involves calculating the momentum sink and TKE source generated by different wind turbines in the wind farm; the second involves distributing the calculated sink and source to the mesoscale grid in order to represent the impact of the wind farm on the mesoscale control equations. Many existing WFP models focus on improving the prediction accuracy of the total momentum sink and TKE source induced by each wind turbine, i.e., the first step. From the classic Fitch model to the explicit wake parametrization (EWP) model that accounts for unresolved wake expansion to the meso-microscale coupled Fitch model that takes subgrid wake effects into account \cite{10fitch2012local,11volker2015explicit,12wu2022refined,13redfern2019incorporation,14pan2018hybrid,15ma2022comparison,16ma2022jensen,17wu2023coupled,18du2025meso}, satisfactory results have been obtained in determinating the total momentum sink and TKE source. However, few studies have focused on their spatial distribution method. The current mainstream spatial distribution method can be referred to as the single-column method. As the name suggests, the influence extent of the sink and source term is limited by a single vertical grid column $(X_i,Y_j)$, which is determined based on the horizontal coordinate of the rotor center. And the distribution weights $W_k$ in different vertical grid layers are equal to the ratio of the intersection area of the rotor swept plane and the vertical grid levels to the rotor area. This method is widely adopted in the classic Fitch model and its numerous extensions\cite{10fitch2012local,12wu2022refined,13redfern2019incorporation,14pan2018hybrid,15ma2022comparison,16ma2022jensen,17wu2023coupled,18du2025meso}. In addition, there is an alternative approach for calculating vertical distribution weights. After the single vertical grid column is determined, the total momentum sink generated by the wind turbine can be distributed across vertical grid levels using a Gaussian vertical profile to account for the unresolved wake expansion within the mesoscale grid. This spatial distribution method is only used in the EWP model proposed by Volker et al. \cite{11volker2015explicit}. In a wind farm, the momentum sink and TKE source generated by the wind farm can be distributed to the mesoscale grid by simply applying the single-column method to each wind turbine. It should be noted that this simple spatial distribution approach is reasonable when the rotor diameter $D$ is small and the mesoscale grid resolution $\Delta$ is coarse. However, it faces significant challenges as turbine sizes increase ($D\sim$200m) and mesoscale simulations become more refined ($\Delta \sim$1 km). As rotor diameter increases, the rotor-swept area also increases significantly. Under these circumstances, the wind turbine is more likely to affect multiple vertical grid columns. However, the conventional single-column method only distributes the sink and source within the single vertical grid column where the rotor center is located, which inevitably leads to misestimations of the spatial distribution of the sink and source.

The main limitation of the single-column approach is that it cannot account for the effects of variations in the rotor-center position within the grid. That is, as long as the turbine’s rotor center lies within the same grid column, the spatial distribution of the generated momentum sink and TKE source are the same. A special case occurs when the turbine is located near a grid intersection. In this case, the rotor-swept area actually covers multiple vertical columns, but the single-column method still assumes that the wind turbine affects only the single vertical grid column where the rotor center is located, and the momentum sink and TKE source are applied only to this grid column. For an entire wind farm, this spatial distribution method will result in aliasing effects in the wind farm layout, leading to an artificial spatial distribution of the sink and source within the wind farm, which may lead to unphysical gaps or jumps, thereby affecting the accuracy of mesoscale flow simulations and may even cause numerical instability. To address these issues, we propose a Gaussian function-based multi-column method to improve the spatial distribution and computational accuracy of the momentum sink and TKE source. Specifically, after calculating the momentum sink and TKE source for each wind turbine, a two-dimensional Gaussian function centered at the turbine rotor center is used to project the sink and source onto multiple vertical grid columns. The projection weights of different grid columns are analytically determined by integrating the two-dimensional Gaussian function over the mesoscale grid. Finally, the vertical distribution weights are further determined using the same approach as the single-column method. By applying the proposed Gaussian-based multi-column method, the sink and source of the mesoscale grid not only depend on the wind turbines within the grid, but also appropriately account for the contributions from wind turbines near grid boundaries, whose rotor centers are located in adjacent grids. This is more consistent with realistic conditions, effectively addressing the misestimation and concentrated distribution of the momentum sink and TKE source inherent in the existing single-column method.

The concept of distributing physical quantities (e.g., thrust, lift, and drag) onto the computational grid using the Gaussian function has been widely used in numerous wind turbine actuator models for large-eddy simulations (LES) \cite{19li2022review}, the actuator wind farm (AWF) model in Reynolds-averaged Navier-Stokes (RANS) simulations \cite{20van2022new,21van2024improved}, and the wind farm model within the meso-micro atmospheric perturbation framework \cite{22allaerts2019sensitivity,23devesse2024meso}. In LES, wind turbines are generally modeled as body forces in microscale computational grids \cite{19li2022review}. The classical actuator disk model treats the wind turbine as a penetrable disk, represented as a finite-thickness cylinder. It is usually necessary to use a Gaussian kernel function to apply the thrust generated by each radial element of the disk to the computational grid \cite{24shapiro2019filtered}. Similarly, in the actuator disk model with rotation, the actuator line model, and the actuator surface model based on blade element theory, the lift and drag generated by any airfoil segment are typically projected onto the three-dimensional computational grid using a Gaussian kernel function \cite{25wu2011large,26martinez2015large,27sorensen2002numerical,28yang2018new}. In RANS simulations, van der Laan et al. \cite{20van2022new,21van2024improved} proposed the actuator wind farm model, investigated the issues caused by the binning method in the AWF model, and proposed a force-distribution scheme for the AWF model using two-dimensional Gaussian functions. In the atmospheric perturbation model for wind farm blockage proposed by Allaerts et al. \cite{22allaerts2019sensitivity} and Devesse et al. \cite{23devesse2024meso}, the turbine thrust is applied to the linearized depth-averaged momentum equations via Gaussian filtering, thereby representing the turbine influence and improving numerical stability. To date, no study has systematically evaluated or compared different spatial distribution methods for the momentum sink and TKE source in WFP models. To the best of the authors' knowledge, this is the first time a two-dimensional Gaussian function has been applied to determine the spatial distribution of the sink and source in WFP models.

The remainder of this paper is structured as follows. Sec. \ref{sec2} details the proposed Gaussian-based multi-column method and its application to the Fitch model. Sec. \ref{sec3} introduces the verification case setups. Simulation results from different WFP models are compared and discussed in Sec. \ref{sec4}. Finally, a summary of the work is presented in Sec. \ref{sec5}.

\section{\label{sec2}The Gaussian-based multi-column method and its application to wind farm parameterizations}
\subsection{\label{sec2.1}The multi-column method based on the two-dimensional Gaussian function}

\begin{figure}
\centering
\includegraphics[width=0.33\textwidth]{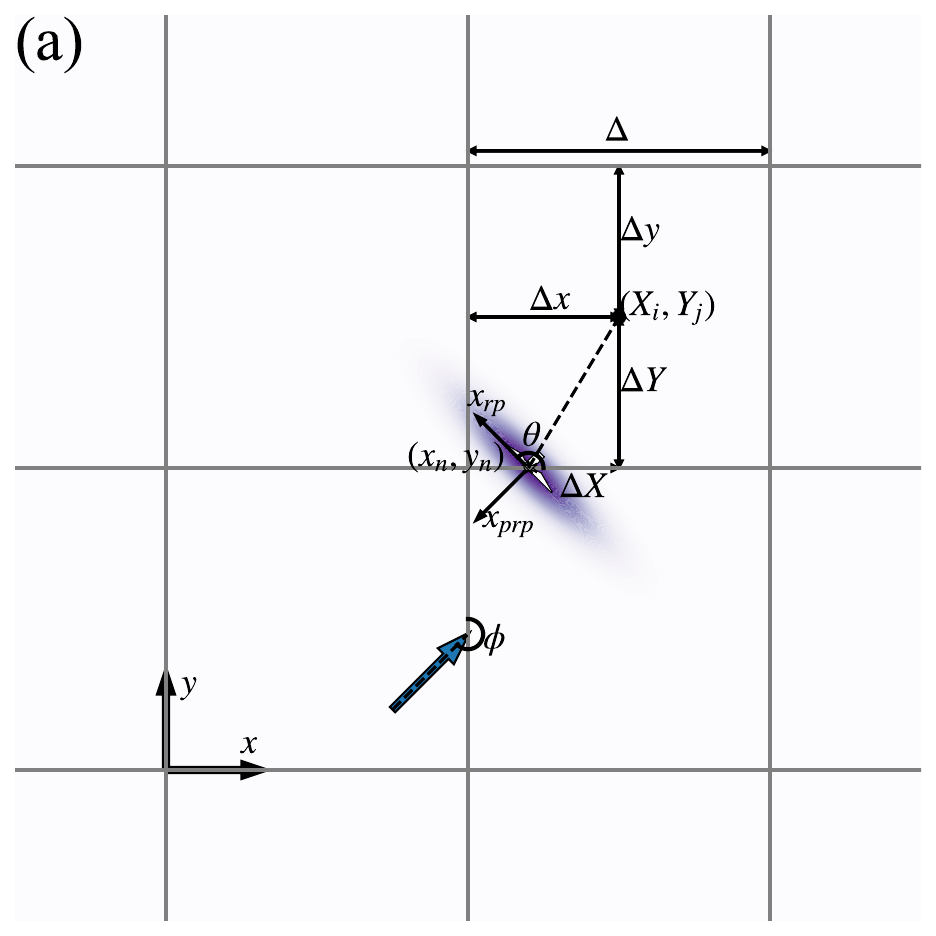}
\includegraphics[width=0.33\textwidth]{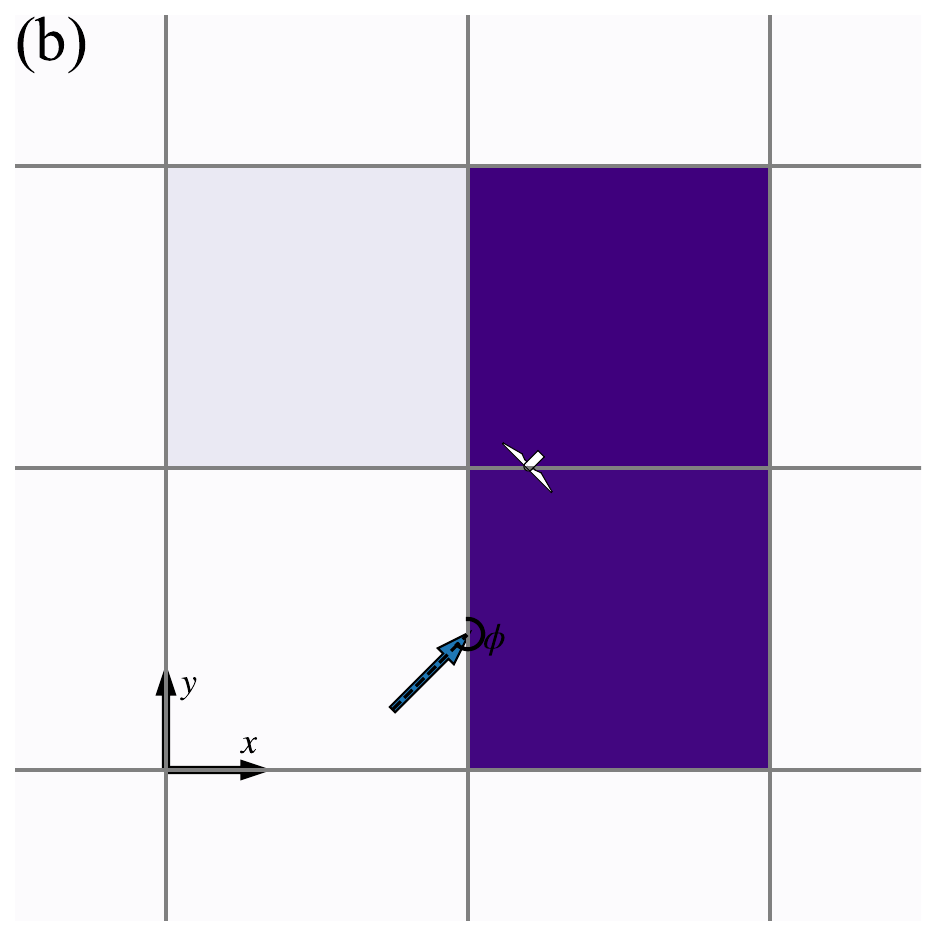}
\includegraphics[width=0.33\textwidth]{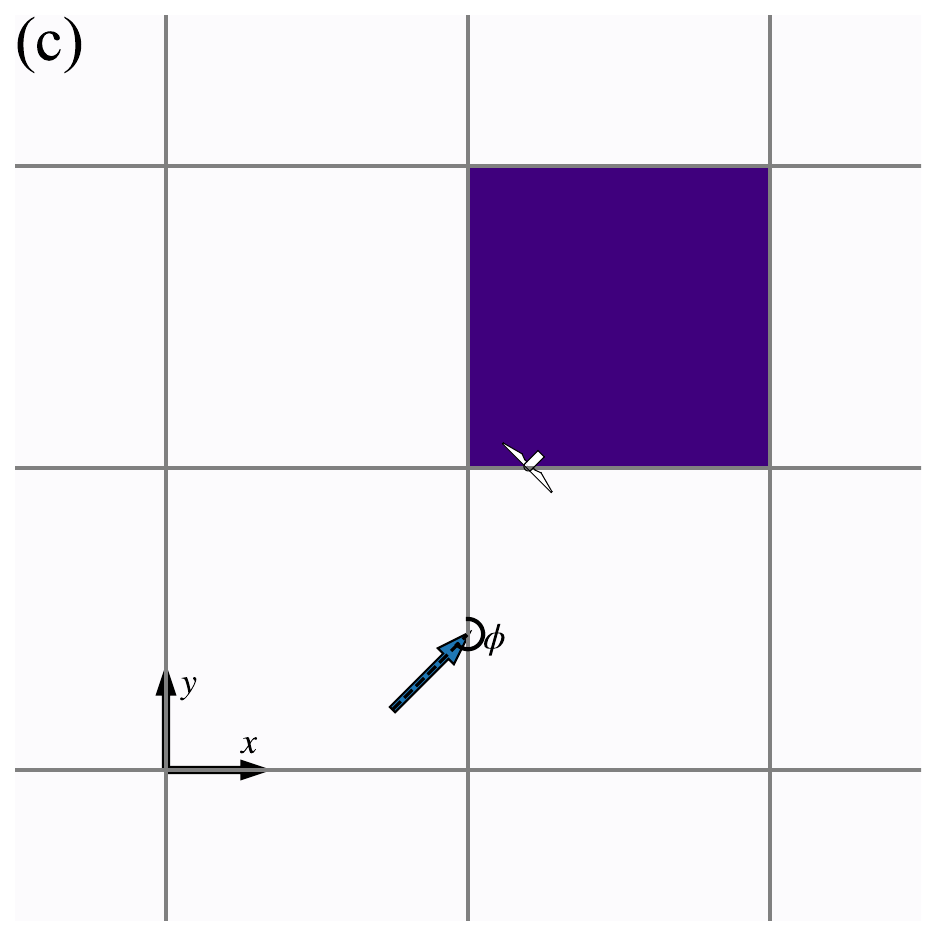}
\caption{\label{fig1} (a) Schematic of calculating the weight $W_{ij,n}$ of the mesoscale vertical grid column $(X_i,Y_j)$ based on the two-dimensional Gaussian function for wind turbine $T_n$, (b) the spatial distribution of $W_{ij,n}$ determined based on the proposed Gaussian-based multi-column method, and (c) the spatial distribution of $W_{ij,n}$ determined based on the conventional single-column method. }
\end{figure}

In this paper, the weights of different mesoscale grid columns are calculated based on the integration of a standard two-dimensional Gaussian function at hub height over the mesoscale grid cells, thus accurately distributing the sink and source generated by each wind turbine onto the computational grid, as shown in figure \ref{fig1}(a). The standard two-dimensional Gaussian function is expressed as follows::
\begin{equation}\label{eq1}
  f_n(\mathbf{x},\mathbf{\mu})=\frac{1}{2\pi \sqrt{\det(\sum)}}\exp\biggl(-\frac{1}{2}(\mathbf{x}-\mathbf{\mu})^T\sum{}^{-1}(\mathbf{x}-\mathbf{\mu})\biggr)
\end{equation}
where $\mathbf{x}=(x,y)$ is an arbitrary point at hub height, $\mathbf{\mu}=(x_n,y_n)$ denotes the coordinates of the rotor center of the wind turbine $T_n$, and $\sum$ is the covariance matrix, which is determined by the projection widths $\sigma_{rp}$ and $\sigma_{prp}$ along and perpendicular to the rotor plane and by the principal-axis direction angle $\theta$, which can be written as:

\begin{equation}\label{eq2}
  \sum=\left(\begin{matrix}
\sigma_{xx} & \sigma_{xy} \\
\sigma_{xy} & \sigma_{yy}
\end{matrix}\right)=\left(\begin{matrix}
\sigma_{rp}^2\cos^2\theta+\sigma_{prp}^2\sin^2\theta & (\sigma_{rp}^2-\sigma_{prp}^2)\sin\theta\cos\theta \\
(\sigma_{rp}^2-\sigma_{prp}^2)\sin\theta\cos\theta & \sigma_{rp}^2\sin^2\theta+\sigma_{prp}^2\cos^2\theta
\end{matrix}\right)
\end{equation}
where the principal-axis angle $\theta=360^\circ-\phi$ is defined as the angle between the rotor plane and the positive direction of the $x$-axis, and $\phi$ denotes the inflow wind direction angle, as shown in figure \ref{fig1}(a).

For each wind turbine $T_n$ located at $(x_n, y_n)$, the weight for the mesoscale grid column $(X_i, Y_j)$ is calculated based on the grid-averaged two-dimensional Gaussian function, and is given by:

\begin{equation}\label{eq3}
  W_{ij,n}=\int_{Y_j-\Delta y}^{Y_j+\Delta y}\int_{X_i-\Delta x}^{X_i+\Delta x}f_n(\mathbf{x},\mathbf{\mu})dx dy
\end{equation}
where $X_i$ and $Y_j$ are the streamwise and spanwise coordinates of the center of the $ij$th mesoscale grid, and $\Delta x=\Delta y=\Delta /2$ are half of the mesoscale grid spacing $\Delta$. Different Gaussian width combinations ($\sigma_{rp}$ and $\sigma_{prp}$) determine the spatial spread of the weighting function. When their values are infinitesimal, the proposed method approaches the single-column method, i.e., $W_{ij,n}=1$ in the grid column containing the rotor center of the wind turbine $T_n$, while $W_{ij,n}=0$ for other grid columns. In the atmospheric perturbation model, Allaerts et al. \cite{22allaerts2019sensitivity} and Devesse et al. \cite{23devesse2024meso} adopted an axisymmetric two-dimensional Gaussian distribution, i.e., by setting $\sigma_{rp}=\sigma_{prp}=\sqrt{2}\Delta$; while in the actuator wind farm model, van der Laan et al. \cite{20van2022new,21van2024improved} used $\sigma_{rp}=2\delta$, $\sigma_{prp}=\max(2\delta, D/4)$ for $(\theta=90^\circ)$, where $D$ is the rotor diameter and $\delta$ is the horizontal grid spacing in the AWF model, which is independent of the mesoscale grid size. Inspired by the above research, we investigate the combinations of different projection widths and finally choose to set $\sigma_{rp}=D$ and  $\sigma_{prp}=D/4$. This combination can achieve a balance between maintaining the geometric characteristics of the wind turbine and ensuring a smooth spatial distribution of the sink and source.

To solve Eq. (\ref{eq3}), a straightforward approach is to divide the mesoscale grid into microscale grid (on the order of 10 m) and perform numerical integration. However, given the large number of wind turbines in a wind farm, each mesoscale grid needs to be finely divided to evaluate the Gaussian function and calculate the total contribution within it for each wind tubrine. This places an additional computational burden on the WFP model in the mesoscale model, which reduces computational efficiency and is unfavorable for practical engineering applications. Therefore, we present an analytical solution inspired by the analytical rotor-averaged deficit across a rectangular disc proposed by Ali et al. \cite{29ali2025direct}.

According to Ali et al. \cite{29ali2025direct}, to analytically evaluate the integral over the intersection of the two-dimensional Gaussian function and the mesoscale grid, Eq. (\ref{eq1}) can first be written as follows:

\begin{equation}\label{eq4}
  f_n(x,y,x_n,y_n)=\frac{1}{2\pi \sigma_{x}\sigma_{y}}\exp\biggl(-\frac{[(x-x_n)+\omega(y-y_n)]^2}{2\sigma_{x}^2}-\frac{[y-y_n]^2}{2\sigma_{y}^2}\biggr)
\end{equation}
where $\omega=-\frac{\sigma_{xy}}{\sigma_{yy}}$ is a parameter describing the overall tilt of the Gaussian function, $\sigma_x=\sqrt{\sigma_{xx}-\omega^2\sigma_{yy}}$ and $\sigma_y=\sqrt{\sigma_{yy}}$ are the streamwise and spanwise Gaussian widths, respectively. By defining $x^\prime=x-x_n$, $y^\prime=y-y_n$, $\Delta X=X_i-x_n$, and $\Delta Y=Y_j-y_n$, Eq. (\ref{eq3}) can then be written as:
\begin{equation}\label{eq5}
  W_{ij,n}=\frac{1}{2\pi \sigma_{x}\sigma_{y}}\int_{\Delta Y-\Delta y}^{\Delta Y+\Delta y}dy^\prime\exp\left(-\frac{{y^\prime}^2}{2\sigma_{y}^2}\right)\times\int_{\Delta X-\Delta x}^{\Delta X+\Delta x}dx^\prime\exp\left(-\frac{(x^\prime+\omega y^\prime)^2}{2\sigma_{x}^2}\right)
\end{equation}

Following Ali et al. \cite{29ali2025direct}, we further define $\sigma=\sigma_{y}$ and $\xi=\sqrt{1-(\sigma_x/\sigma_y)^2}(\sigma_y\ge\sigma_x)$. Finally, we obtain the analytical expression for $W_{ij,n}$, as follows:
\begin{equation}\label{eq6}
  W_{ij,n}=\sum_{s_x,s_y\in[-1,1]}(-s_xs_y)\times\Omega^*\left(\frac{\Delta Y+s_y\Delta y}{\sigma},\frac{\omega}{\sqrt{1-\xi^2}},\frac{\Delta X+s_x\Delta x}{\sigma\sqrt{1-\xi^2}}\right)
\end{equation}
If $\sigma_y<\sigma_x$, we define $\sigma=\sigma_x$ and $\xi=\sqrt{1-(\sigma_y/\sigma_x)^2}$, and $W_{ij,n}$ is be expressed as:
\begin{equation}\label{eq7}
  W_{ij,n}=\sum_{s_x,s_y\in[-1,1]}(-s_xs_y)\times\Omega^*\left(\frac{\Delta Y+s_y\Delta y}{\sigma\sqrt{1-\xi^2}},\omega\sqrt{1-\xi^2},\frac{\Delta X+s_x\Delta x}{\sigma}\right)
\end{equation}
where $\Omega^*(h,a,b)$ is:
\begin{equation}\label{eq8}
  \Omega^*(h,a,b)=-\frac{1}{2\pi}\left(\arctan(a+b/h)+\arctan\left(\frac{h+ab+a^2h}{b}\right)\right)+T(h,a+b/h)+T\left(\frac{b}{\sqrt{1+a^2}},\frac{h+ab+a^2h}{b}\right)
\end{equation}
where $T(h,a)$ is Owen’s T function defined as \cite{30owen1956tables}:
\begin{equation}\label{eq9}
  T(h,a)=\frac{1}{2\pi}\int_0^a\frac{1}{1+s^2}\exp\left(\frac{-h^2(1+s^2)}{2}\right)ds
\end{equation}
It should be noted that only the final expressions of $W_{ij,n}$ are given here. The detailed derivation can be found in Ali et al. \cite{29ali2025direct}.

Figures \ref{fig1}(b) and (c) further illustrate the grid weights obtained from the Gaussian-based multi-column method and the conventional single-column method. As can be seen, the proposed method avoids the shortcomings of the single-column approach and effectively captures the influence of wind turbines on multiple grid columns. Furthermore, the analytical expression reduces computational burden and can be easily integrated into existing WFP models.

Theoretically, determining grid weights by integrating the two-dimensional Gaussian function is consistent with calculating them from the intersection area of the rotor plane with multiple vertical grid columns. However, given the complexity of calculating the intersection area between the rotor plane and multiple vertical grid columns and the difficulty of formulating a generalized analytical solution, we adopt the former in the paper. For multi-column horizontal grid weights at different height levels, we assume them to be approximately identical to the horizontal weight distribution at hub height. This assumption significantly reduce the additional computational cost, allowing the grid weight distribution at different height levels to be obtained with a single calculation, while keeping the introduced error within an acceptable range. To further reduce the computational cost, it is unnecessary to evaluate the weights of every grid cell in the mesoscale computational domain; instead, only the weights within the region of influence of the wind farm $\Omega_{wf}=\{(i,j)|i\in I, j\in J\}$ are considered.  $I=\{i|i_{s}-1<=i<=i_{e}+1\}$ and $J=\{j|j_s-1<=j<=j_e+1\}$ are the streamwise and spanwise index set, where $i_s$ and $i_e$ ($j_s$ and $j_e$) are the minimum and maximum values of the streamwise (spanwise) grid index in which the wind turbine is located.

\subsection{\label{sec2.2}Implementation of the proposed method in the WFP models}
To demonstrate the advantages of the proposed method, we integrated the Gaussian-based multi-column spatial distribution method into the classic Fitch model \cite{10fitch2012local}, and implemented it within the open-source mesoscale Weather Research and Forecasting (WRF) model \cite{31skamarock2019description}. The Fitch model has been integrated into numerous mesoscale models and now serves as the default WFP in the WRF system, which has been widely applied and validated in the literature. Detailed computational procedures are described in Fitch et al. \cite{10fitch2012local}. After calculating the momentum sink and TKE source for each wind turbine, we added an additional procedure to compute the grid-column weights using the proposed Gaussian-based multi-column method, according to the relative positions of the mesoscale grids and turbine rotor centers. Subsequently, the sink and source terms were spatially distributed based on these weights. On this basis, we developed a refined WFP model and defined it as the Fitch-Gaussian model. The expressions for the momentum sink and TKE source in these two WFP models are summarized in table \ref{tab1}.
\begin{table}
\centering
\renewcommand{\arraystretch}{1.25}
\caption{Expressions for the momentum sink and TKE source in the Fitch and Fitch-Gaussian model}\label{tab1}
\begin{tabular*}{1.00\linewidth}{@{\extracolsep{\fill}}lccc@{}}
\toprule
\makecell[l]{WFP \\model} & \makecell[c]{Spatial \\distribution \\method} & $u-$ ($v$-)momentum sink & TKE source \\
\midrule
Fitch & \makecell[c]{Single-\\column}&
\makecell[c]{
 $\left(\frac{\partial u}{\partial t}\right)_{wf}=\sum_{n\in N_{ij}}-\frac{1}{2}\frac{1}{A_{cell}}\frac{A_{k,n}C_{T,n}U_{ijk}u_{ijk}}{(z_{k+1}-z_k)}$ \\
 $\left(\frac{\partial v}{\partial t}\right)_{wf}=\sum_{n\in N_{ij}}-\frac{1}{2}\frac{1}{A_{cell}}\frac{A_{k,n}C_{T,n}U_{ijk}v_{ijk}}{(z_{k+1}-z_k)}$}
&
\makecell[c]{$\left(\frac{\partial TKE}{\partial t}\right)_{wf}=\sum_{n\in N_{ij}}\frac{1}{2}\frac{1}{A_{cell}}\frac{A_{k,n}C_{TKE,n}U_{ijk}^3}{(z_{k+1}-z_k)}$}
\\
\makecell[l]{Fitch-\\Gaussian} & \makecell[c]{Gaussian-\\based \\multi-\\column}&
\makecell[c]{
  $\left(\frac{\partial u}{\partial t}\right)_{wf}=\sum_{n=1}^{N}-\frac{1}{2}\frac{1}{A_{cell}}\frac{A_{k,n}C_{T,n}\mathfrak{U}_{k,n}\mathfrak{u}_{k,n}}{(z_{k+1}-z_k)}W_{ij,n}$  \\
  $\left(\frac{\partial v}{\partial t}\right)_{wf}=\sum_{n=1}^{N}-\frac{1}{2}\frac{1}{A_{cell}}\frac{A_{k,n}C_{T,n}\mathfrak{U}_{k,n}\mathfrak{v}_{k,n}}{(z_{k+1}-z_k)}W_{ij,n}$}
&
\makecell[c]{$\left(\frac{\partial TKE}{\partial t}\right)_{wf}=\sum_{n=1}^{N}\frac{1}{2}\frac{1}{A_{cell}}\frac{A_{k,n}C_{TKE,n}\mathfrak{U}_{k,n}^3}{(z_{k+1}-z_k)}W_{ij,n}$}
\\
\bottomrule
\end{tabular*}
\end{table}

In table \ref{tab1}, $\left(\frac{\partial u}{\partial t}\right)_{wf}$ and $\left(\frac{\partial v}{\partial t}\right)_{wf}$ are the $u$- and $v$- momentum sinks induced by the wind farm, respectively. $\left(\frac{\partial TKE}{\partial t}\right)_{wf}$ represents the TKE source induced by the wind farm. They are directly integrated into the mesoscale governing equations of momentum and TKE. In their expressions, $N_{ij}$ is the turbine index set associated with the vertical grid column $ij$, $N$ is the number of turbines in the wind farm, $k$ is the vertical index of the vertical column, $A_{cell}=\Delta^2$ is the mesoscale grid-cell area, $z_{k+1}$ and $z_k$ are the vertical coordinates at levels $k+1$ and $k$, and $A_{k,n}$ is the intersection area of the rotor swept area with the vertical grid layers between levels $k$ and $k+1$. $A_{k,n}W_{ij,n}$ can be regarded as the intersection area of the wind turbine $T_n$ with grid cell $ijk$. $C_{T,n}$ and $C_{P,n}$ are the thrust and power coefficients of wind turbine $T_n$, respectively. $\alpha$ is an empirical correction coefficient, and $C_{TKE,n}=\alpha(C_{T,n}-C_{P,n})$ represents the TKE source correction factor. $u_{ijk}$, $v_{ijk}$, and $U_{ijk}=\sqrt{u_{ijk}^2+u_{ijk}^2}$  are the streamwise, spanwise, and horizontal wind velocities at grid cell $ijk$, respectively. Correspondingly, $\mathfrak{u}_{k,n}=\sum_{i\in I}\sum_{j \in J}u_{ijk}W_{ij,n}$, $\mathfrak{v}_{k,n}=\sum_{i\in I}\sum_{j \in J}v_{ijk}W_{ij,n}$, and $\mathfrak{U}_{k,n}=\sum_{i\in I}\sum_{j \in J}U_{ijk}W_{ij,n}$ are the weighted streamwise, spanwise, and horizontal wind velocities corresponding to each wind turbine $T_n$ at different vertical grid levels. By replacing the grid wind speed of the Fitch model with the weighted wind speed and considering the spatial distribution weights $W_{ij,n}$ of each wind turbine, the corresponding momentum sink and TKE source expressions for the Fitch-Gaussian model are obtained, as shown in table \ref{tab1}.

Essentially, the $u$- ($v$-)momentum sink in the mesoscale model directly corresponds to the projection of the turbine thrust in streamwise (spanwise) direction, whereas the physical meaning of the TKE source is more complex. As theoretically derived by Abkar et al. \cite{32abkar2015new}, the TKE source generated by the wind farm in the mesoscale model equals the rate of work done by velocity fluctuations against the turbine-induced forces in the spatially averaged TKE budget equation, which can be written as follows:

\begin{equation}\label{eq10}
  P_{TKE}=-\left[\sum_{n\in N_{ij}}\left(\frac{\partial u}{\partial t}\right)_n(U_{ijk}-U_{d,n})\frac{u_{ijk}}{U_{ijk}}+\sum_{n\in N_{ij}}\left(\frac{\partial v}{\partial t}\right)_n(U_{ijk}-U_{d,n})\frac{v_{ijk}}{U_{ijk}}\right]
\end{equation}
where $U_{d,n}$ is the rotor-equivalent wind speed of wind turbine $T_n$. Substituting the momentum sink expressions from the Fitch model into the above expression, and using the one-dimensional momentum theory $C_{T,n}=4a(1-a)$, $C_{P,n}=4a(1-a)^2$ and $U_{d,n}=U_{ijk}(1-a)$, we obtain:
\begin{equation}\label{eq11}
  P_{TKE}=\sum_{n\in N_{ij}}\frac{1}{2}\frac{1}{A_{cell}}\frac{A_{k,n}C_{T,n}U_{ijk}^2}{(z_{k+1}-z_k)}(U_{ijk}-U_{d,n})=\sum_{n\in N_{ij}}\frac{1}{2}\frac{1}{A_{cell}}\frac{A_{k,n}(C_{T,n}-C_{P,n})U_{ijk}^3}{(z_{k+1}-z_k)}
\end{equation}
However, it has been found that directly using the above expression leads to an overestimation of the TKE downstream of the wind turbine \cite{33archer2020two}. Therefore, it is suggested that an empirical correction coefficient $\alpha$ be introduced to resolve this issue. In this case, Eq. (\ref{eq11}) becomes identical to the TKE source expression of the Fitch model, as shown in table \ref{tab1}.
\begin{equation}\label{eq12}
  P_{TKE}=\sum_{n\in N_{ij}}\frac{1}{2}\frac{1}{A_{cell}}\frac{A_{k,n}\alpha(C_{T,n}-C_{P,n})U_{ijk}^3}{(z_{k+1}-z_k)}
\end{equation}

Based on the LES results for a stand-alone wind turbine case, Archer et al. \cite{33archer2020two} conducted a parameter sensitivity analysis and showed that $\alpha=0.25$ is an optimal parameter, leading to results that are closer to the TKE obtained in mesoscale simulations. However, practical experience in wind farm simulations suggests that a higher empirical coefficient is needed. LESs of wind farms indicate that $\alpha= 0.5$ is a more suitable coefficient \cite{18du2025meso,34radunz2025under}, while some comparisons against field measurements indicate that $\alpha= 1.0$ provides more accurate predictions \cite{35larsen2021case,36ali2023assessment}. To date, the determination of $\alpha$ remains an open question. Since the main focus of this paper is to propose and verify the advantages of the proposed spatial distribution method, providing a unified method for determining this coefficient is beyond the scope of this paper and will be addressed in future work. Therefore, we set $\alpha=0.5$ in the case setups of this paper, as this value ensures a similar distribution of the total TKE source in mesoscale simulations and LESs.

Finally, it is worth noting that the proposed method is generally applicable to the Fitch model and its numerous refined versions, but it is incompatible with the EWP model. This is because the EWP model assumes that the rotor center is located at the mesoscale grid center during its derivation, which contradicts our basic assumption of determining grid weights based on the relative position of the mesoscale grid and the rotor center.

\section{\label{sec3}Verification case setups}
\begin{figure}
\centering
\includegraphics[width=0.9\textwidth]{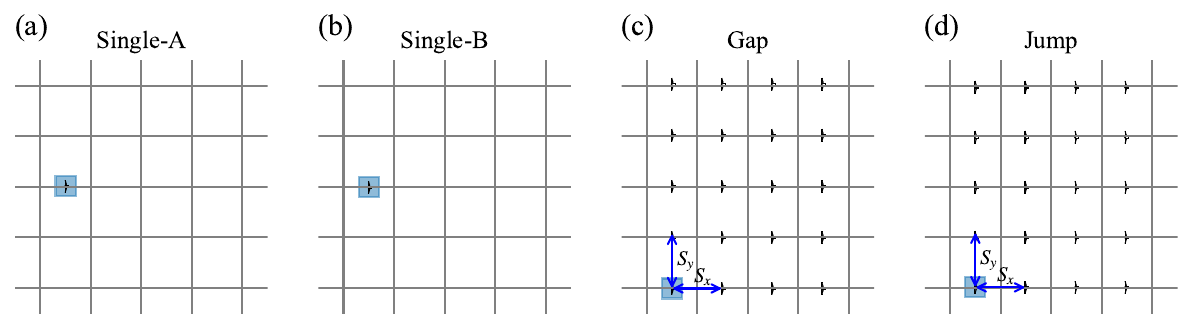}
\caption{\label{fig2} Schematic of the wind farm layouts and their relative locations with the mesoscale grid ($\Delta$=1km). (a) Single-A, (b) Single-B, (c) Gap, and (d) Jump wind farms. (The wind turbine enclosed with a skyblue box is the reference wind turbine.)}
\end{figure}

Based on the high-fidelity LES data of wind farms, we set up verification cases to study the performance of the conventional single-column method and the Gaussian-based multi-column method in characterizing the spatial distribution of the sink and source, and in simulating the evolution of wind-farm wakes. LES is generally used as the benchmark in the verification of WFP model improvements \cite{32abkar2015new,33archer2020two,34radunz2025under,37pena2022evaluation,38garcia2023evaluation,39garcia2024mesoscale,40sanchez2025differences}. As introduced in Sec. \ref{sec1}, the proposed method is mainly aimed at addressing the limitations faced by the single-column method when the wind turbine is located at the grid intersection. When the wind turbine is located near the center of the grid, the proposed method is nearly equivalent to the single-column method. Therefore, in the verification cases, we only considered the case where the wind turbine is located at the grid intersection. Specifically, we studied two stand-alone wind turbine cases and two aligned small wind farm cases. The simulated wind farm layouts and their relative locations in the mesoscale grid ($\Delta$=1km) are shown in figure \ref{fig2}. The simulated wind turbine is the IEA10 MW wind turbine with a hub height $H$=119m and a rotor diameter $D$ =198m, with performance curves available in Bortolotti et al. \cite{41bortolotti2019iea}. In the two stand-alone wind turbine cases, the rotor centers were located 10 m above and 10 m below the mesoscale grid boundary, respectively, and were thus named Single-A and Single-B. In the two aligned small wind farm cases, the spanwise spacing $S_y$ of the wind turbines is $5D$ and $5.1D$, respectively. Resulting from the use of the conventional single-column method, an abrupt gap and jump in the spanwise spatial distribution of the sink and source would occur, which are therefore called Gap and Jump, respectively. Detailed parameters of the wind farm layouts in different cases are shown in table \ref{tab2}.
\begin{table}
\renewcommand{\arraystretch}{1.25}
\caption{Overview of the wind farm layouts}\label{tab2}
\begin{tabular*}{0.9\linewidth}{@{\extracolsep{\fill}}lccccc@{}}
\toprule
Case & \makecell[c]{Number of\\ wind turbines} & \makecell[c]{Streamwise \\spacing \\($S_x/D$)} & \makecell[c]{Spanwise \\spacing \\($S_y/D$)} & $(x_{PBL}^r, y_{PBL}^r)$&$(x_{LES}^r, y_{LES}^r)$\\
\midrule
Single-A & $1\times 1$&N/A &N/A&(35500m, 15010m)& (5500m, 5010m)\\
Single-B & $1\times 1$&N/A &N/A&(35500m, 14990m)& (5500m, 4990m)\\
Gap & $4\times 5$&5 &5.1&(35500m, 12990m)& (5500m, 2990m)\\
Jump & $4\times 5$&5 &5&(35500m, 13011m)& (5500m, 3011m)\\
\bottomrule
\end{tabular*}
\end{table}

\begin{figure}
\centering
\includegraphics[width=0.9\textwidth]{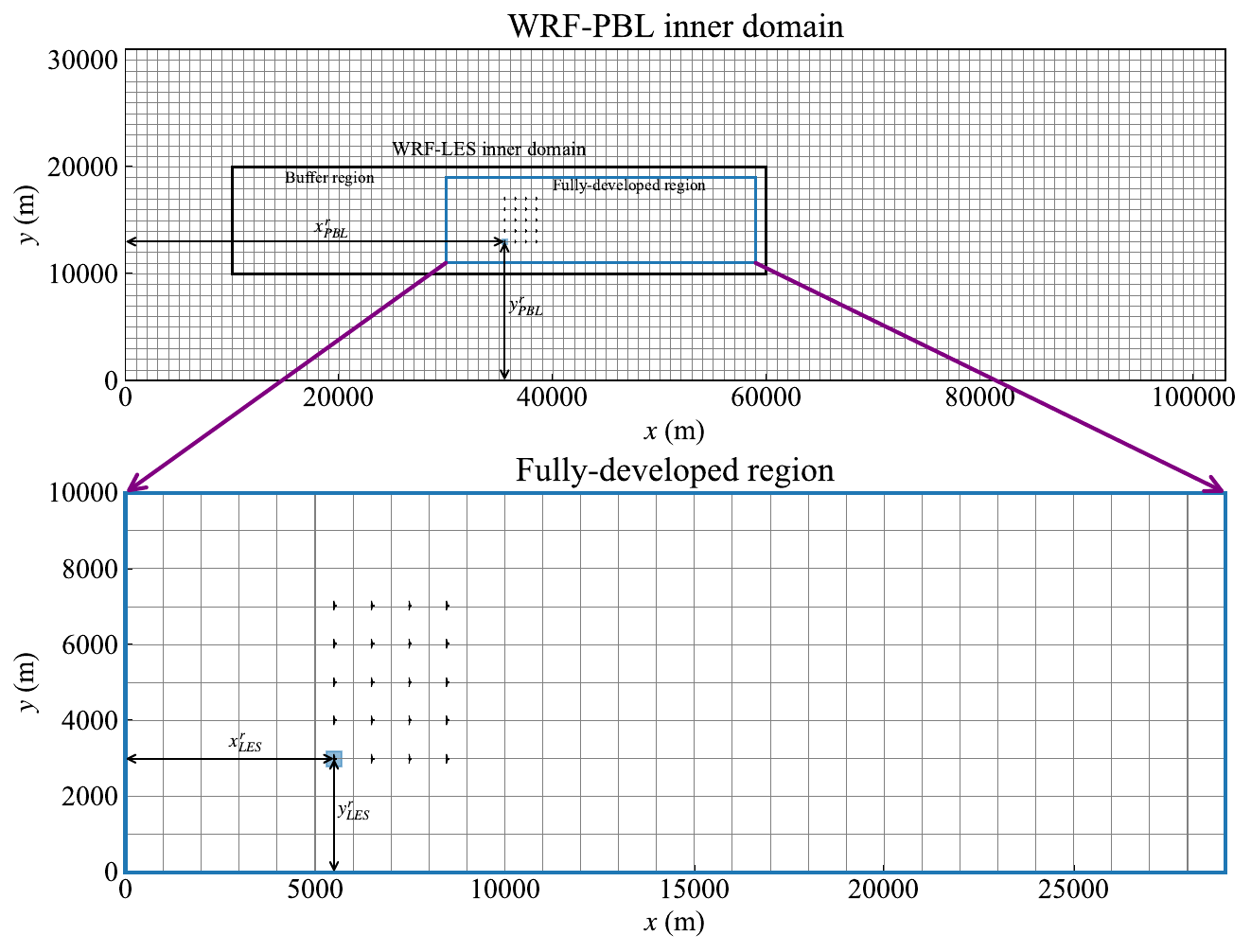}
\caption{\label{fig3} Schematic of the WRF-PBL and WRF-LES inner domain with the Gap wind farm. ($(x_{PBL}^r, y_{PBL}^r)$ and $(x_{LES}^r, y_{LES}^r)$ are the coordinates of the reference wind turbine in the WRF-PBL and WRF-LES inner domain, whose values can be found in table \ref{tab2}.)}
\end{figure}

Both the high-fidelity LES and mesoscale simulation of wind farms are based on the open-source WRF model, though they use different frameworks. The large eddy simulation framework (WRF-LES) uses the Deardorff subgrid model, while the mesoscale simulation framework (WRF-PBL) employs the MYNN-2.5 planetary boundary layer parameterization scheme. Using the same set of codes ensures the consistency of the numerical schemes, thereby reducing numerical differences between large eddy simualtions and mesoscale simulations. This helps focus on the differences caused by different spatial distribution methods \cite{37pena2022evaluation}. The WRF model has integrated a variety of wind farm parameterizations and wind turbine actuator models, and is widely used for simulating wind-farm flows. Similar to García-Santiago et al. \cite{38garcia2023evaluation}, we ran the WRF model in idealized mode with minimal physical parameterizations to isolate the development of the atmospheric boundary layer within the WRF-LES and WRF-PBL frameworks. In both frameworks, two domains with one-way nesting are adopted, with periodic boundary conditions for the outer lateral boundaries and the nested boundary conditions for the inner lateral boundaries. We use the same surface-layer scheme in which Monin–Obukhov similarity theory is applied. In the mesoscale WRF-PBL framework, a uniform horizontal grid $103 \times 31$ is used, with an outer horizontal grid resolution of 3 km and an inner horizontal grid resolution of 1 km. A non-uniform vertical grid is used, with a total of 120 vertical levels, the model top at 2 km, and the lowest 40 model levels spaced at 5 m intervals. Considering the computational cost, wind farm layout, and the nature of turbulent flow development at the transition region between the inner and outer domains of the WRF-LES framework, a coarse WRF-LES computation grid with sufficient streamwise length is used. A uniform horizontal grid $1000 \times 202$ is used, with an outer horizontal grid resolution of 150 m and an inner horizontal grid resolution of 50 m. The vertical discretization is the same as that of the WRF-PBL framework. To avoid the influence of boundary conditions on the WRF-LES inner domain, a buffer region is introduced in the inner domain. On the western boundary, the buffer region is 20 km long, and the cell perturbation method (CPM) \cite{42munoz2014bridging,43munoz2015stochastic} is used to ensure that the simulated wind farm is immersed in fully-developed turbulence. On other boundaries, the buffer regions are 1 km long, as shown in figure \ref{fig3}.

\begin{figure}
\includegraphics[width=1.0\textwidth]{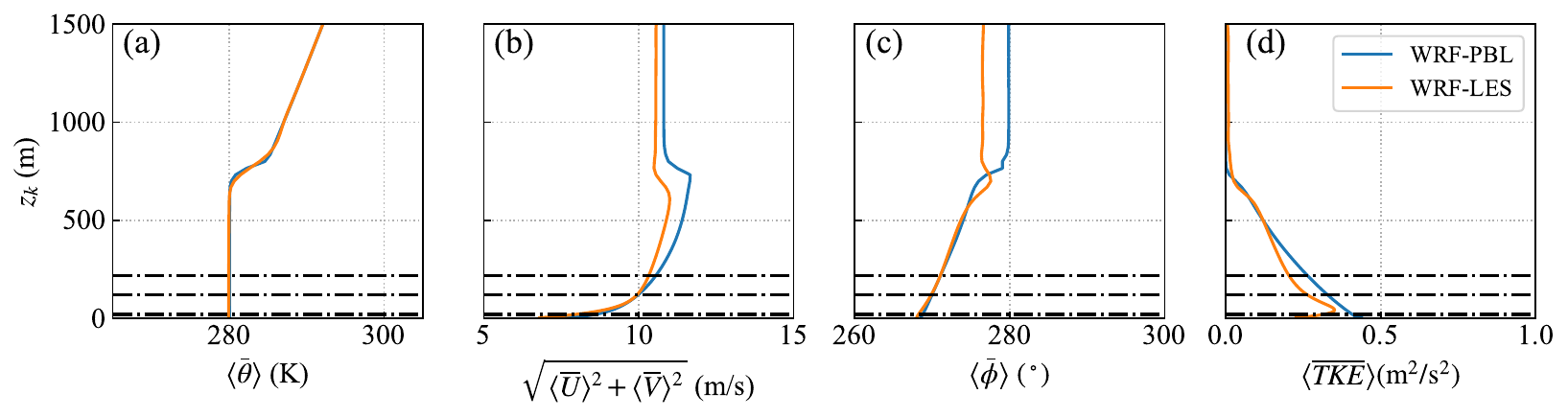}
\caption{\label{fig4} Comparison of the last-hour and spatial averaged vertical inflow profiles corresponding to the fully developed region of the WRF-LES framework and the inner domain of the WRF-PBL framework. (a) Potential temperature, (b) horizontal wind speed, (c) wind direction, and (d) turbulent kinetic energy. The top, bottom, and hub height of the wind turbine rotor are depicted with horizontal black lines for reference.}
\end{figure}

To fairly compare the simulation results of the WRF-LES and WRF-PBL frameworks, it is necessary to ensure that the inflow wind conditions generated by both frameworks are as similar as possible. Following the work of García-Santiago et al. \cite{38garcia2023evaluation}, we set a typical conventional neutral boundary layer condition in the North Sea region of Europe. The surface roughness was set to 0.0002m (a typical offshore roughness value), and the Coriolis force parameter was set to $1.14\times10^{-4}s^{-1}$, corresponding to $51.6^\circ$N. The initial potential temperature profile is characterized by an inversion layer height of 700 m, an inversion layer thickness of 100 m, an inversion layer temperature difference of 5K, and a potential temperature lapse rate of 10K/km in the free atmospheric boundary layer, as shown in figure \ref{fig4}. By adjusting the magnitude of the geostrophic wind components in the WRF-LES and WRF-PBL frameworks, we can ensure that the hub-height-averaged wind direction is about $270^\circ$ and the average wind speed is about 10 m/s after reaching the quasi-steady state. The case setups of the WRF-LES and WRF-PBL frameworks are shown in table \ref{tab3}.

\begin{table}
\centering
\renewcommand{\arraystretch}{1.25}
\caption{Overview of the case setups of the WRF-LES and WRF-PBL framworks}\label{tab3}
\begin{tabular*}{1.0\linewidth}{@{\extracolsep{\fill}}lcccccccc@{}}
\toprule
Framework & Domain & \makecell[c]{Horizontal \\grid \\$(N_x\times N_y)$} & \makecell[c]{Horizontal\\ grid resolution \\(m)}&\makecell[c]{Time \\step \\(s)}&\makecell[c]{Output \\frequency\\ (s)}& \makecell[c]{Geostrophic\\ wind component\\ $G_x$, $G_y$(m/s)}&\makecell[c]{Spin-up \\time (h)}&\makecell[c]{Total \\run \\time (h)}\\
\midrule
\multirow{2}{*}{WRF-PBL}& Outer& \multirow{2}{*}{$103\times31$} & 3000 &30 &- &\multirow{2}{*}{10.65, -1.85} &\multirow{2}{*}{18} &\multirow{2}{*}{24}\\
 & Inner&  & 1000 & 10 & 600& &  &\\
\multirow{2}{*}{WRF-LES}& Outer& \multirow{2}{*}{$1000\times202$} & 150 &1.5 &- &\multirow{2}{*}{10.47, -1.19} &\multirow{2}{*}{16} &\multirow{2}{*}{18}\\
 & Inner&  & 50 & 0.5 & 10& &  &\\
\bottomrule
\end{tabular*}
\end{table}
Figure \ref{fig4} compares the last hour and spatial averaged vertical inflow profiles for the fully-developed region of the WRF-LES framework and the inner domain of the WRF-PBL framework. It can be seen that, except for a slight deviation of the TKE profiles, our case setups ensure that the inflow conditions within the wind turbine rotor region are in good agreement, providing a solid foundation for subsequent comparisons.

After the spin-up time, we can set up the corresponding wind turbine actuator model or WFP model in different frameworks and restart the simulation to model the interaction between wind farm and ABL. For each wind farm layout shown in table \ref{tab2}, we carried out microscale large-eddy simulation and mesoscale simulation. In the WRF-LES framework, the SADLES model proposed by Bui et al. \cite{44bui2024implementation} was used to model the wind turbine. This method is consistent with the uniform actuator disk model, where the body force of the wind turbine is distributed according to the intersection area of the wind rotor swept area and the vertical grid. Previous studies have confirmed that this type of actuator model can still well capture the time-averaged flow of wind farms under a coarse 50 m grid \cite{44bui2024implementation,45stipa2024actuator}, which satisfactorily meets our requirements. In the WRF-PBL framework, we carried out mesoscale simulations with the Fitch model using the conventional single-column method and the Fitch-Gaussian model using the proposed Gaussian-based multi-column model to compare and highlight the advantages of the proposed spatial distribution method. Taking the Gap wind farm as an example, the relative positions of wind turbines in the inner domain of WRF-PBL and WRF-LES frameworks are depicted in figure \ref{fig3}. The positions of the reference wind turbines for different wind farms in the inner domain can be found in table \ref{tab2}.

In view of the flow transition induced by wind turbines, we selected quantities of interest averaged over the last hour of the total run time for subsequent comparison. To directly compare the results of WRF-PBL with those of WRF-LES, we spatially averaged the time-averaged quantities obtained from the WRF-LES framework using the grid size of the mesoscale inner domain (i.e., 1 km$\times$1 km) to match the spatial resolution of different numerical frameworks. The corresponding quantities can be expressed as $\langle q \rangle_{\Delta}$, where $q$ is a quantity of interest in the microscale WRF-LES framework. Unless otherwise specified, the quantities in the following WRF-LES cases are the results of spatial averaged over the mesoscale grid, and $\langle\cdot\rangle_{\Delta}$ is omitted for simplicity. Furthermore, because the inner domain sizes of WRF-LES and WRF-PBL differ, the coordinates of the spatial averaged results obtained from the WRF-LES framework are transformed according to the mesoscale grid of WRF-PBL framework, allowing for direct comparison of simulation results within the same computational grid.

\section{\label{sec4}Results and discussion}
\subsection{\label{sec41} Comparison of the momentum sink and TKE source}
The momentum sink and TKE source are direct outputs of WFP models and are integrated into the mesoscale governing equations to influence mesoscale flows. Therefore, with LES results as a benchmark, we first compare the magnitude and spatial distribution of the momentum sink and TKE source obtained from different WFP models. Actually, the $u$- ($v$-) momentum sink in the mesoscale model corresponds to the spatial averaged streamwise (spanwise) body force on the mesoscale grid in the WRF-LES framework \cite{14pan2018hybrid,32abkar2015new}. For simplicity, we compare the magnitude of the total momentum sink $T$ induced by the wind turbine in the WRF-LES framework, that is:
\begin{figure}
\centering
\includegraphics[width=0.8\textwidth]{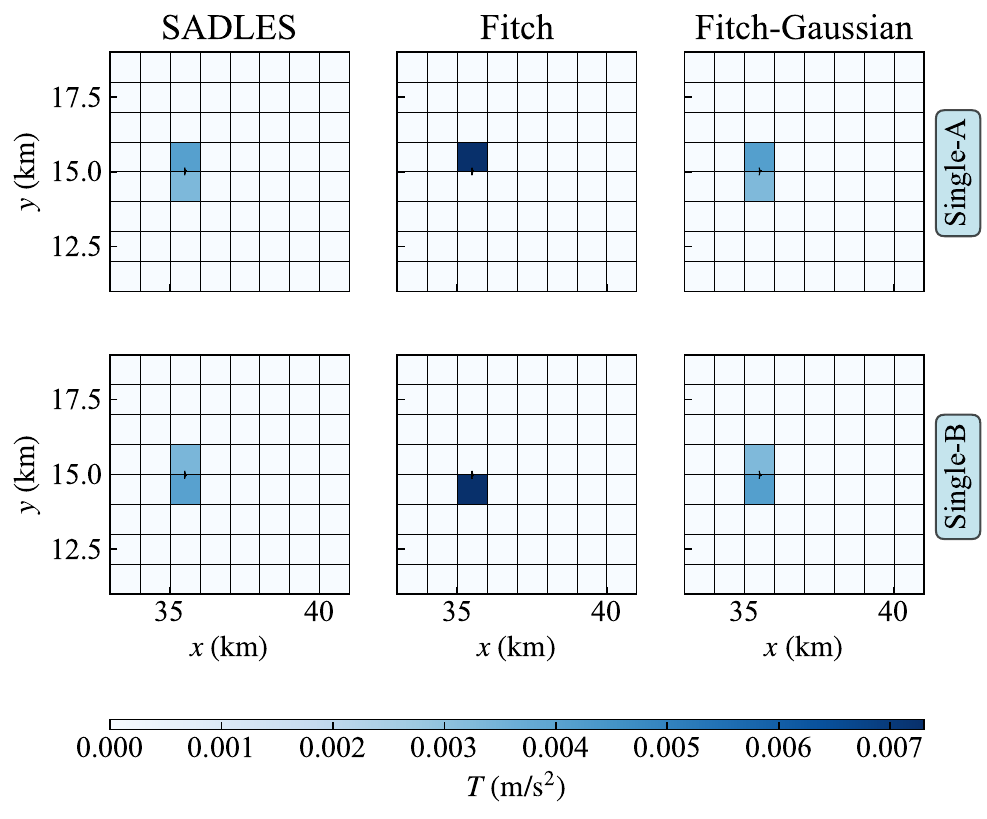}
\caption{\label{fig5} Comparison of the hub-height spatial distribution of the total momentum sink $T$ for the Single-A and Single-B cases obtained from the WRF-LES and WRF-PBL frameworks}
\end{figure}
\begin{figure}
\centering
\includegraphics[width=0.8\textwidth]{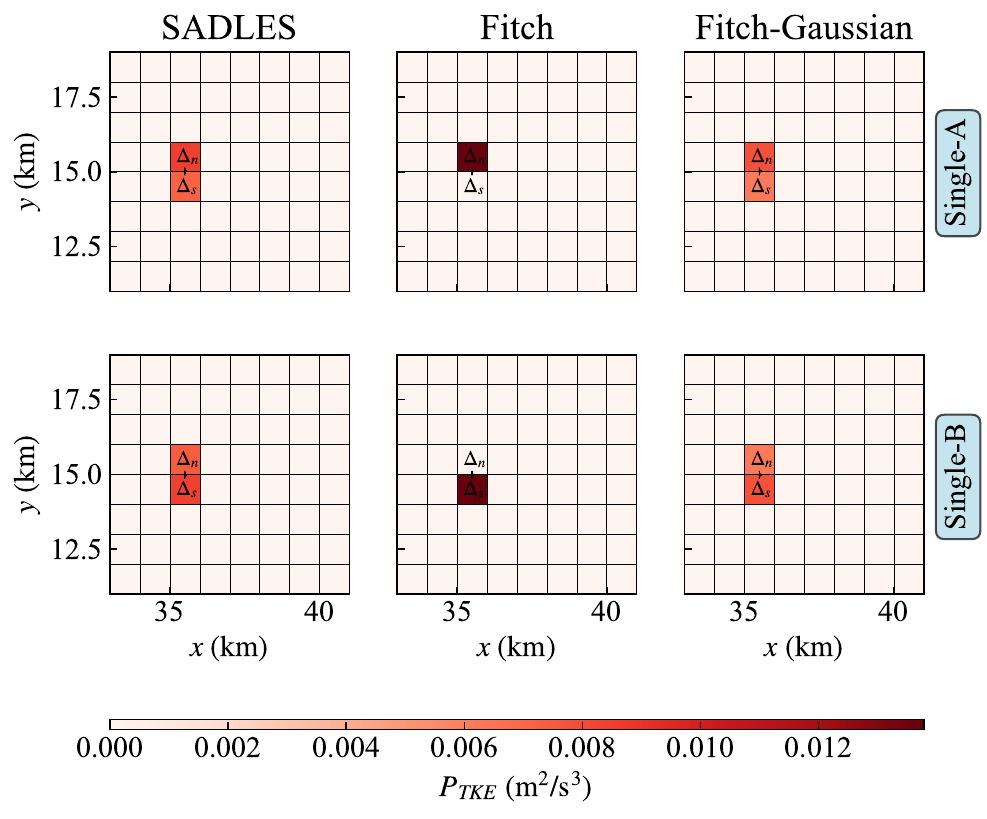}
\caption{\label{fig6} Comparison of the hub-height spatial distribution of the TKE source $P_{TKE}$ for the Single-A and Single-B cases obtained from the WRF-LES and WRF-PBL frameworks ($\Delta_n$ and $\Delta_s$ are the labels of vertical grid columns.)}
\end{figure}
\begin{figure}
\centering
\includegraphics[width=0.9\textwidth]{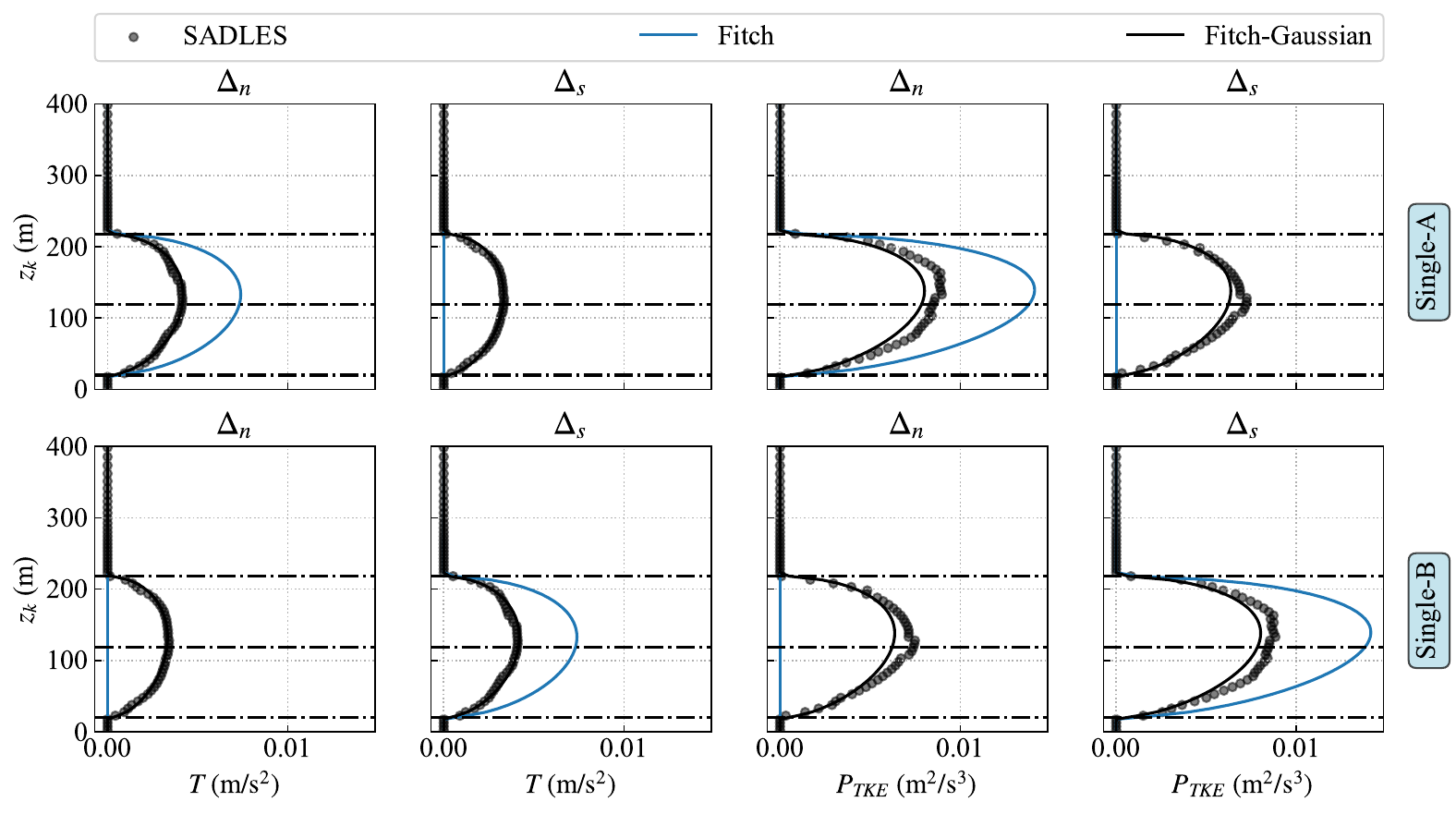}
\caption{\label{fig7} Comparison of the the vertical profiles of the total momentum sink $T$ and TKE source $P_{TKE}$ in the vertical grid columns $\Delta_n$ and $\Delta_s$ for the Single-A and Single-B cases obtained from the WRF-LES and WRF-PBL frameworks. The top, bottom, and hub height of the wind turbine rotor are depicted with horizontal black lines for reference.}
\end{figure}
\begin{equation}\label{eq13}
  T=\sqrt{\langle \overline{f_u}\rangle_{\Delta}^2+\langle \overline{f_v}\rangle_{\Delta}^2}
\end{equation}
where $\overline{\cdot}$ is the temporal average, $\langle\cdot\rangle_\Delta$ represents the spatial average over the mesoscale grid, and $f_u$ and $f_v$ are the instantaneous streamwise and spanwise body force, which can be extracted directly from the output of the wind turbine SADLES model.

The TKE source in the WRF-LES framework refers to the correlation between velocity and body force spatial fluctuation, which can be expressed as \cite{14pan2018hybrid}:
\begin{equation}\label{eq14}
  P_{TKE}=-\langle \overline{u_i^{\prime\prime}f_i^{\prime \prime}}\rangle_\Delta=-(\langle\overline{u_if_i}\rangle_\Delta-\langle\overline{u_i}\rangle_\Delta\langle\overline{f_i}\rangle_\Delta)
\end{equation}
where $\cdot^{\prime\prime}=\overline{\cdot}-\langle\overline{\cdot}\rangle_\Delta$ represents the deviation from the mesoscale grid average, and $u_i$ is the instantaneous velocity in the $i$-direction, which can be extracted from the output of the WRF-LES framework. Therefore, the reference momentum sink and TKE source of the WRF-LES framework can be easily obtained after simple post-processing.

\subsubsection{\label{sec411} Stand-alone wind turbine cases}
Figures \ref{fig5} and \ref{fig6} show the hub-height spatial distribution of the total momentum sink and TKE source for the Single-A and Single-B cases obtained from WRF-LES and WRF-PBL frameworks, respectively. As expected, the sink and source calculated by the single-column method are concentrated within the grid where the rotor center is located, resulting in a localized distribution within a single grid. In comparison, the Fitch-Gaussian model more accurately predicts the impact of the wind turbine on the surrounding mesoscale grid and has spatial distributions closer to those obtained with SADLES.

Figure \ref{fig7} further shows the vertical profiles of the total momentum sink and TKE source in the grid columns $\Delta_n$ and $\Delta_s$ for the Single-A and Single-B cases obtained from WRF-LES and WRF-PBL frameworks. It can be seen that the sum of the sink and source generated by the Fitch and Fitch-Gaussian model is nearly the same, but there are significant differences in their spatial distribution. The sink and source predicted by the Fitch model are distributed in the single vertical grid column where the rotor center is located, significantly overestimating the sink and source within this grid column and underestimating them in the surrounding vertical grid columns. In contrast, the proposed Gaussian-based multi-column method can effectively address this problem. The vertical distributions of the sink and source within different vertical grid columns for the Fitch-Gaussian model are nearly identical to the results obtained from WRF-LES framework.
\begin{figure}
\centering
\includegraphics[width=0.79\textwidth]{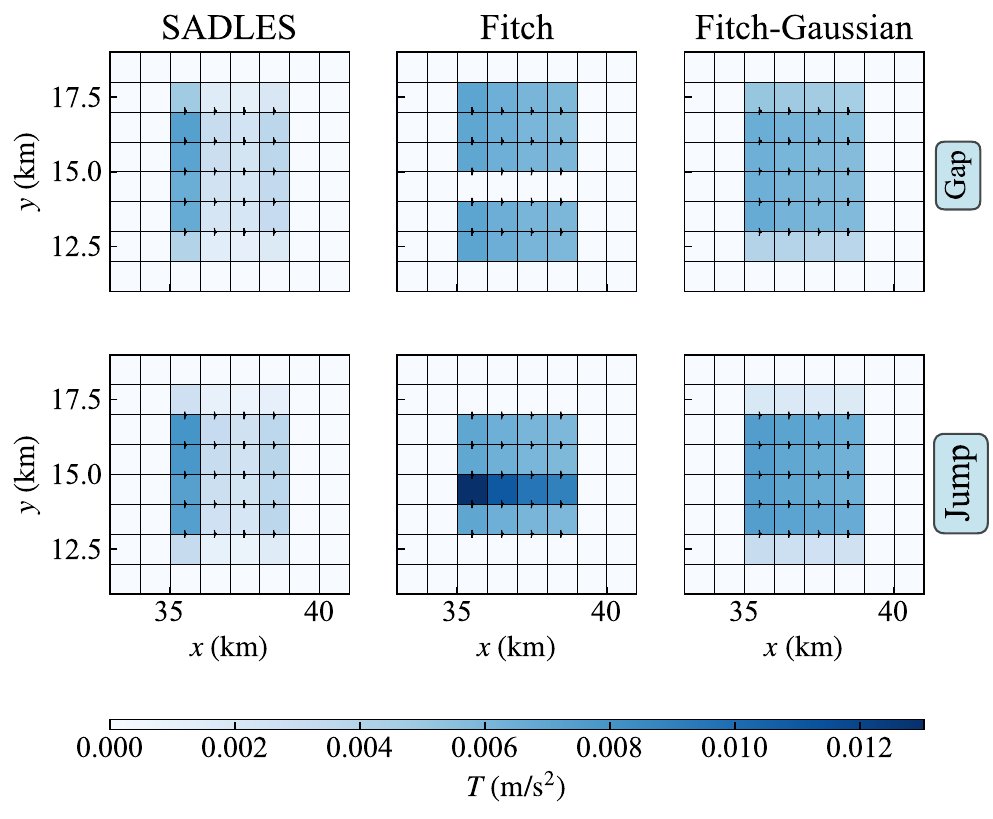}
\caption{\label{fig8} Comparison of the hub-height spatial distribution of the total momentum sink $T$ for the Gap and Jump cases obtained from the WRF-LES and WRF-PBL frameworks}
\end{figure}

\begin{figure}
\centering
\includegraphics[width=0.8\textwidth]{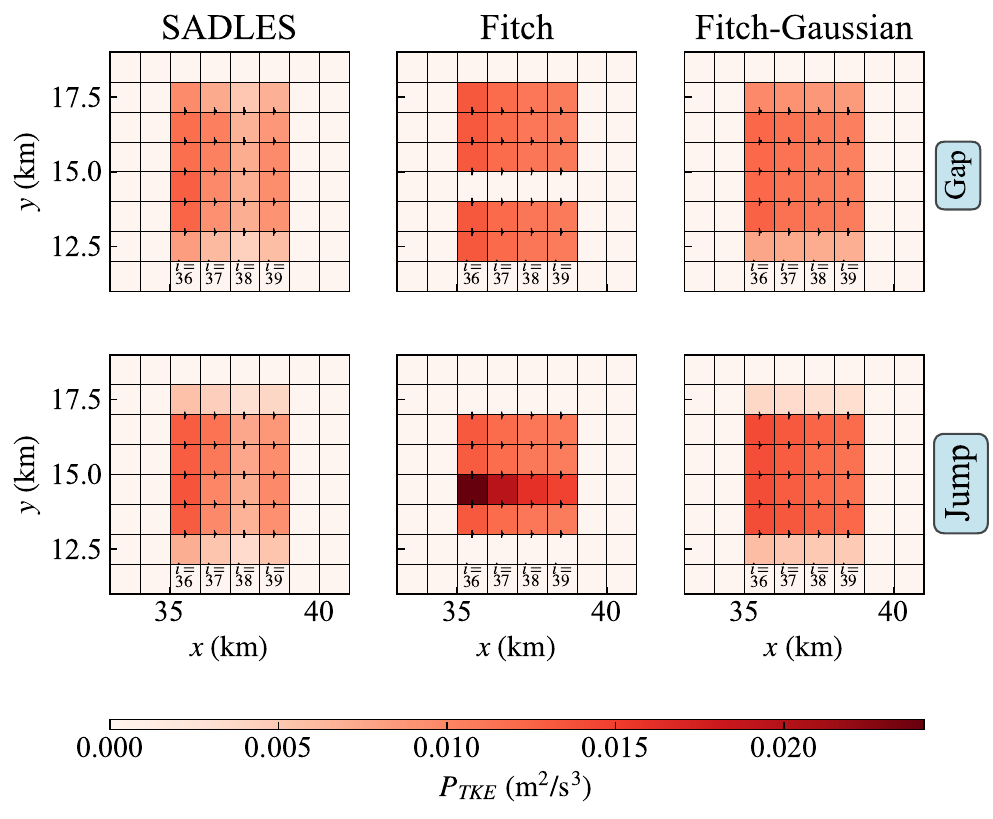}
\caption{\label{fig9} Comparison of the hub-height spatial distribution of the TKE source $P_{TKE}$ for the Gap and Jump cases obtained from the WRF-LES and WRF-PBL frameworks}
\end{figure}
\subsubsection{\label{sec412} Wind farm cases}
Figures \ref{fig8} and \ref{fig9} show the hub-height spatial distribution of the total momentum sink and TKE source for the Gap and Jump cases obtained from WRF-LES and WRF-PBL frameworks, respectively. It can be seen that although the two simulated wind farm layouts are homogeneous in the spanwise direction, the Fitch model exhibits an abrupt jump (gap) in the spatial distribution of the sink and source within the Jump (Gap) wind farm due to the single-column method, resulting in an artificially aliased wind farm layout. In contrast, the spatial distribution of the sink and source using the Gaussian-based multi-column method is suitable for these conditions and maintains the spanwise homogeneity within the Jump (Gap) wind farm, effectively capturing the essential characteristics of the wind farm layout and more closely resembling the spatial distributions of the sink and source obtained from WRF-LES framework. Nevertheless, the streamwise spatial distribution details predicted by the Fitch-Gaussian model show some difference from the LES results.

\begin{figure}
\centering
\includegraphics[width=0.9\textwidth]{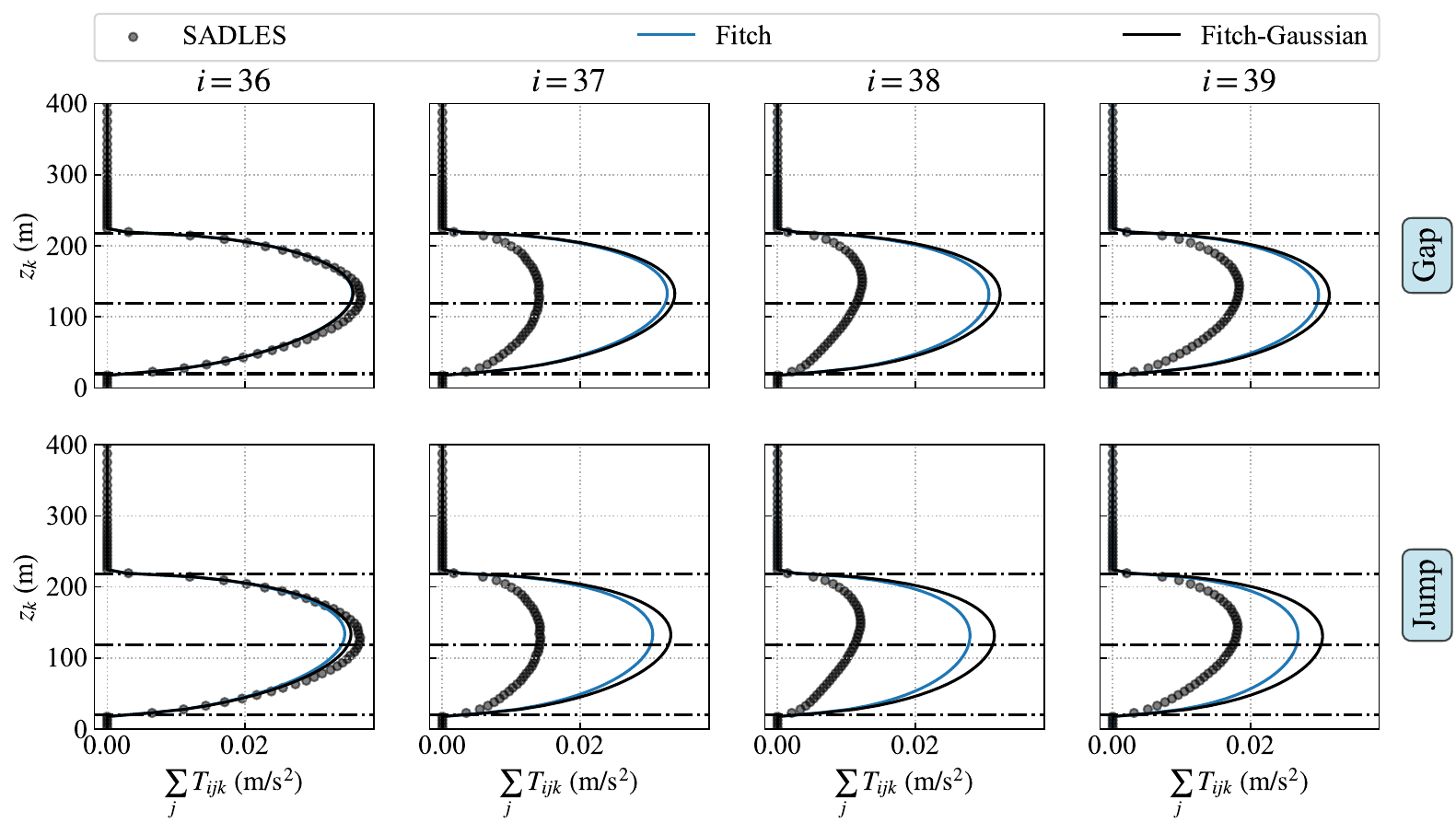}
\caption{\label{fig10} Comparison of the vertical spatial distribution of the sum of the total momentum sink at different streamwise grid indices within the wind farm footprint  (i.e. from $i=36$ to $i=39$ in figure \ref{fig9}) obtained from the WRF-LES and WRF-PBL frameworks. The top, bottom, and hub height of the wind turbine rotor are depicted with horizontal black lines for reference.}
\end{figure}

\begin{figure}
\centering
\includegraphics[width=0.9\textwidth]{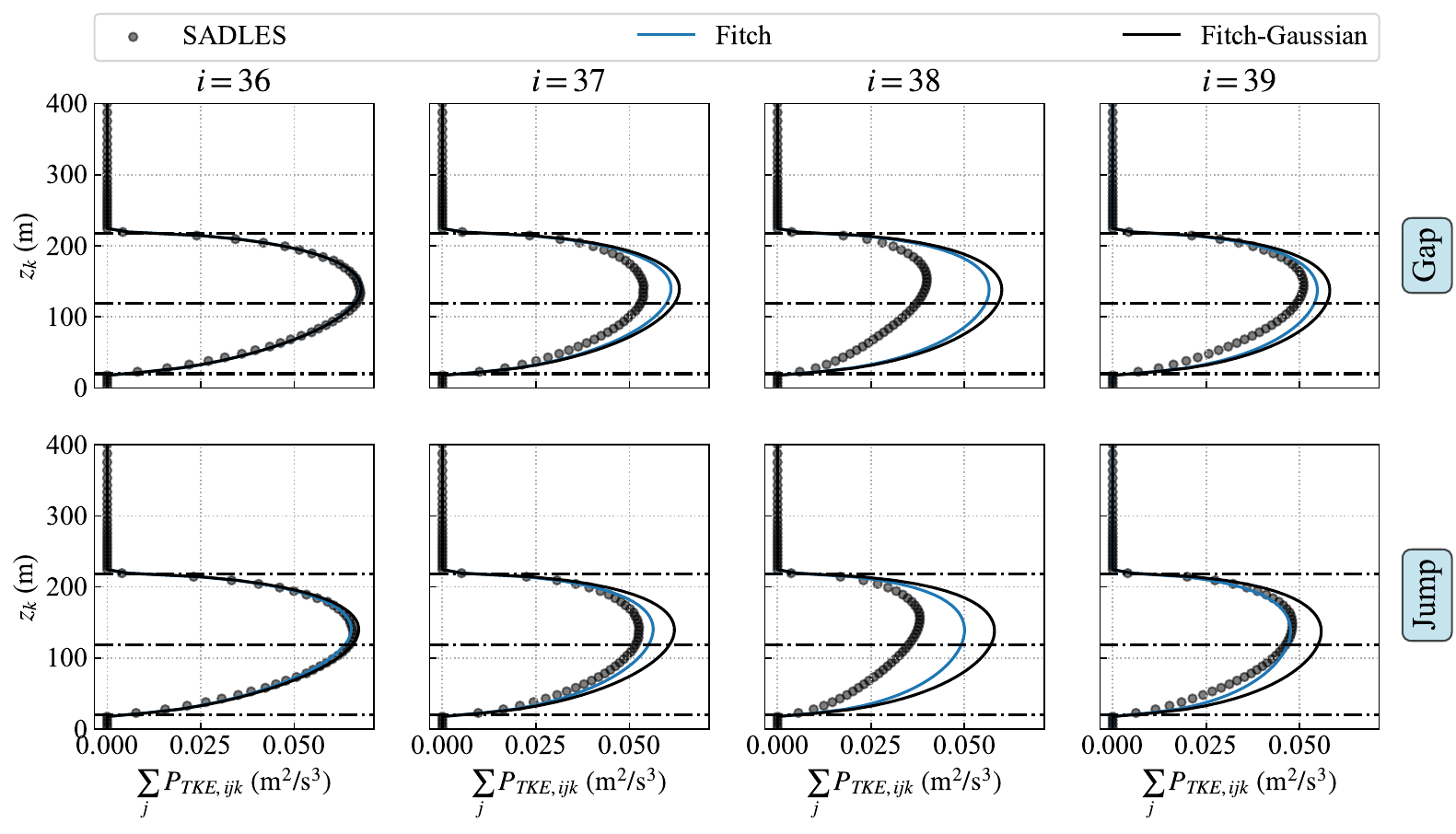}
\caption{\label{fig11} Comparison of the vertical spatial distribution of the sum of the TKE source at different streamwise grid indices within the wind farm footprint  (i.e. from $i=36$ to $i=39$ in figure \ref{fig9}) obtained from the WRF-LES and WRF-PBL frameworks. The top, bottom, and hub height of the wind turbine rotor are depicted with horizontal black lines for reference.}
\end{figure}

To further illustrate this, figures \ref{fig10} and \ref{fig11} further show the vertical spatial distribution of the sum of the total momentum sink and TKE source at different streamwise grid indices within the wind farm footprint (i.e. from $i=36$ to $i=39$ in figure \ref{fig9}) obtained from WRF-LES and WRF-PBL frameworks. It can be seen that the sink and source predicted by the Fitch and the Fitch-Gaussian model behave very similarly, demonstrating a gradual decrease trend downstream. This is mainly because the sink and source in these two WFP models are determined from mesoscale grid wind speeds. The mesoscale grid velocities only account for convection from upstream grid wind speed and turbine-induced velocity deficit at the mesoscale scale and cannot fully resolve the wake interactions and wake recovery, which are much smaller than the mesoscale grid size, resulting in a gradual decrease in mesoscale grid wind speeds within wind farms. This differs significantly from the conditions in LESs. In LESs, the thrust (i.e., the momentum sink) is determined based on the rotor-equivalent inflow wind speed. In the entrance region of the aligned wind farm, the inflow wind speed of the second-row wind turbines decreases significantly due to the wake of the first row, while the inflows of further downstream wind turbines are influenced by the combined wake deficits and added turbulence; they tend to reach a plateau and even have a slight increase. Therefore, the corresponding momentum sink predicted by LES initially decreases abruptly and then slightly increases within the wind farm. Similarly, the TKE source shows a similar trend, but the initial decrease is significantly weaker than the momentum sink. Overall, the TKE source predicted by the Fitch and Fitch-Gaussian model shows satisfactory agreement with that of LES results, indicating the effectiveness of setting the empirical correction coefficient $\alpha$ as 0.5.

Recently, Du et al. \cite{18du2025meso} proposed a meso-microscale coupled wind farm parameterization (MMC-Fitch) model, coupling a microscale wind-farm flow analytical model with the mesoscale WRF model to calculate the local inflow wind speed of the wind turbine. In the future, the proposed Gaussian-based multi-column method an be combined with the MMC-Fitch model to further improve the streamwise spatial distribution details of the sink and source.

\subsubsection{\label{sec413} Statistical metrics}
To further quantify the prediction errors of the total momentum sink and TKE source for different WFP models, we define the spatial averaged normalized root mean square error ($NRMSE$) and spatial distribution correlation coefficient ($R$) based on the WRF-LES results:
\begin{equation}\label{eq15}
  NRMSE_{\eta}=\frac{\sqrt{\frac{1}{V_{\neq0}}\iiint_{V_{\neq0}}\left(\eta^{PBL}-\eta^{LES}\right)^2dV}}{\max(\eta^{LES})}
\end{equation}

\begin{equation}\label{eq16}
  R_\eta=\frac{\iiint_{V_{\neq0}}(\eta^{PBL}_{ijk}-\overline{\eta^{PBL}_{ijk}})(\eta^{LES}_{ijk}-\overline{\eta^{LES}_{ijk}})dV}{\sqrt{\iiint_{V_{\neq0}}(\eta^{PBL}_{ijk}-\overline{\eta^{PBL}_{ijk}})^2dV}\sqrt{\iiint_{V_{\neq0}}(\eta^{LES}_{ijk}-\overline{\eta^{LES}_{ijk}})^2dV}}
\end{equation}
where $V_{\neq0}$ corresponds to the volume where the sink and source are non-zero, $\eta$ is the physical quantity of interest (i.e., $T$ and $P_{TKE}$), and the superscripts $^{PBL}$ and $^{LES}$ denote the quantities from the WRF-PBL and WRF-LES frameworks, respectively. Additionally, $\overline{...}$ represents the spatial average over $V_{\neq0}$. For example, $\overline{\eta^{PBL}_{ijk}}=\frac{1}{V_{\neq 0}}\iiint_{V_{\neq0}}\eta^{PBL}_{ijk}dV$.
\begin{table}
\centering
\renewcommand{\arraystretch}{1.25}
\caption{The spatial averaged normalized root mean square error of the total momentum sink and TKE source for the Fitch and Fitch-Gaussian model}\label{tab4}
\begin{tabular*}{0.7\linewidth}{@{\extracolsep{\fill}}lcccc@{}}
\toprule
\multirow{2}{*}{Wind farm} & \multicolumn{2}{c}{Fitch} & \multicolumn{2}{c}{Fitch-Gaussian}  \\
 & $NRMSE_T$ & $NRMSE_{P_{TKE}}$ & $NRMSE_T$ &$NRMSE_{P_{TKE}}$ \\
\midrule
Single-A & 0.64 & 0.53 & 0.03 & 0.08  \\
Single-B & 0.66 & 0.56 & 0.03 & 0.08 \\
Gap      & 0.40 & 0.34 & 0.29 & 0.14  \\
Jump     & 0.36 & 0.61 & 0.28 & 0.33  \\
\midrule
Mean    & 0.52 & 0.51 & 0.16 & 0.16  \\
\bottomrule
\end{tabular*}
\end{table}

\begin{table}
\centering
\renewcommand{\arraystretch}{1.25}
\caption{The spatial distribution correlation coefficient of the total momentum sink and TKE source for the Fitch and Fitch-Gaussian model}\label{tab5}
\begin{tabular*}{0.7\linewidth}{@{\extracolsep{\fill}}lcccc@{}}
\toprule
\multirow{2}{*}{Wind farm} & \multicolumn{2}{c}{Fitch} & \multicolumn{2}{c}{Fitch-Gaussian} \\
                           & $R_T$ & $R_{P_{TKE}}$ & $R_T$ & $R_{P_{TKE}}$ \\
\midrule
Single-A & 0.59 & 0.64 & 0.99 & 0.99 \\
Single-B & 0.55 & 0.59 & 0.99 & 0.99 \\
Gap      & 0.26 & 0.37 & 0.67 & 0.91 \\
Jump     & 0.67 & 0.82 & 0.74 & 0.92 \\
\midrule
Mean     & 0.52 & 0.61 & 0.85 & 0.95 \\
\bottomrule
\end{tabular*}
\end{table}
Tables \ref{tab4} and \ref{tab5} show the normalized root mean square errors and correlation coefficients for the momentum sink and TKE source for different WFP models, respectively. Overall, compared with the Fitch model, the Fitch-Gaussian model using the Gaussian-based multi-column method significantly improves the spatial distribution correlation coefficient and reduce the $NRMSE$ of the momentum sink and TKE source. In the stand-alone wind turbine cases, the wind turbine is unaffected by the wake of the upstream turbine, and the mesoscale grid wind speed is very close to the rotor equivalent inflow wind speed. The performance of the Fitch-Gaussian model is significantly better than the Fitch model. In the wind farm cases, the rotor-equivalent inflow wind speed of the downstream wind turbines in the wind farm differs significantly from the mesoscale grid wind speed due to the influence of wind-turbine wakes, and the prediction error of both the Fitch and Fitch-Gaussian model increases slightly. Nevertheless, the Fitch-Gaussian model is still better than the Fitch model. The averaged $NRMSE_T$ and $NRMSE_{P_{TKE}}$ for all four wind farms decrease from 0.52 and 0.51 to 0.16 and 0.16. The averaged $R_T$ and $R_{P_{TKE}}$ for all four wind farms increase from 0.52 and 0.61 to 0.85 and 0.95. It is worth noting that the total momentum sink and TKE source calculated by the two WFP models are very close (see figures \ref{fig7}, \ref{fig10} and \ref{fig11}), indicating that the improvement in the statistical metrics is entirely due to the proposed spatial distribution method, which further demonstrates the advantages of the Gaussian-based multi-column method.
\subsection{\label{sec42}Comparison of the evolution of wind-farm wakes}
After analyzing the magnitude and spatial distribution of the momentum sink and TKE source, we proceed to analyze the simulated wind-turbine wakes. To quantify the wake evolution, we define the normalized velocity deficit $\lambda_{\Delta U}$ and the normalized added TKE $\lambda_{\Delta TKE}$ of the wind-farm wakes as follows \cite{36ali2023assessment}:
\begin{equation}\label{eq15}
  \lambda_{\Delta U}=\frac{\Delta U}{U_{NT}}=\frac{U-U_{NT}}{U_{NT}} \qquad \lambda_{\Delta TKE}=\frac{\Delta TKE}{TKE_{NT}}=\frac{TKE-TKE_{NT}}{TKE_{NT}}
\end{equation}
where $U\ (TKE)$ is the streamwise wind speed (turbulent kinetic energy) from the WRF-LES or WRF-PBL framework with wind turbines, $U_{NT}$ ($TKE_{NT}$) represents the streamwise wind speed (turbulent kinetic energy) from the WRF-LES or WRF-PBL framework without wind turbines, and $\Delta U$ and $\Delta TKE$ are the velocity deficit and added TKE, respectively.
\subsubsection{\label{sec421} Comparison of velocity deficit}
\begin{figure}
\centering
\includegraphics[width=0.9\textwidth]{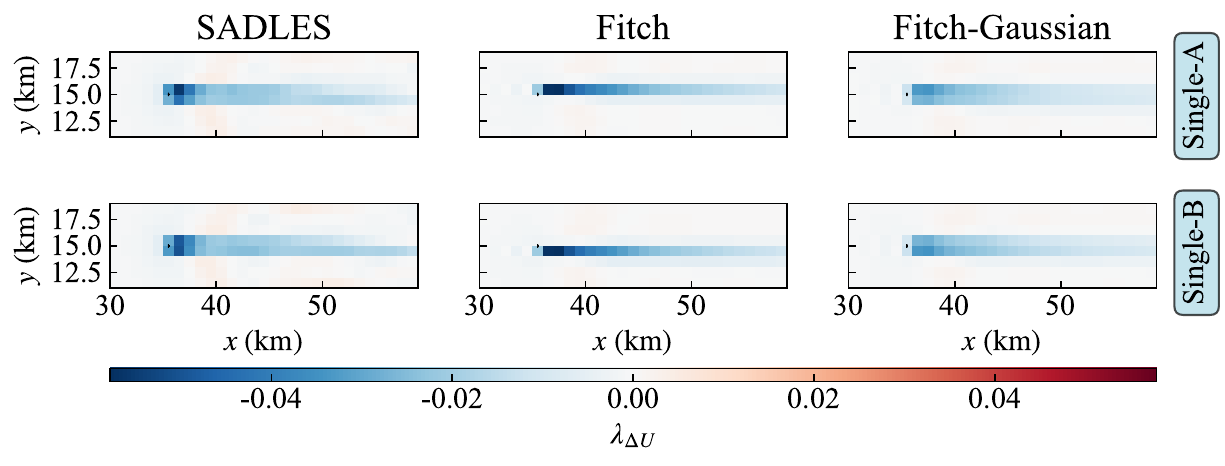}
\caption{\label{fig12} Comparison of the hub-height normalized velocity deficits for Single-A and Single-B cases obtained from the WRF-LES and WRF-PBL frameworks}
\end{figure}

\begin{figure}
\centering
\includegraphics[width=0.9\textwidth]{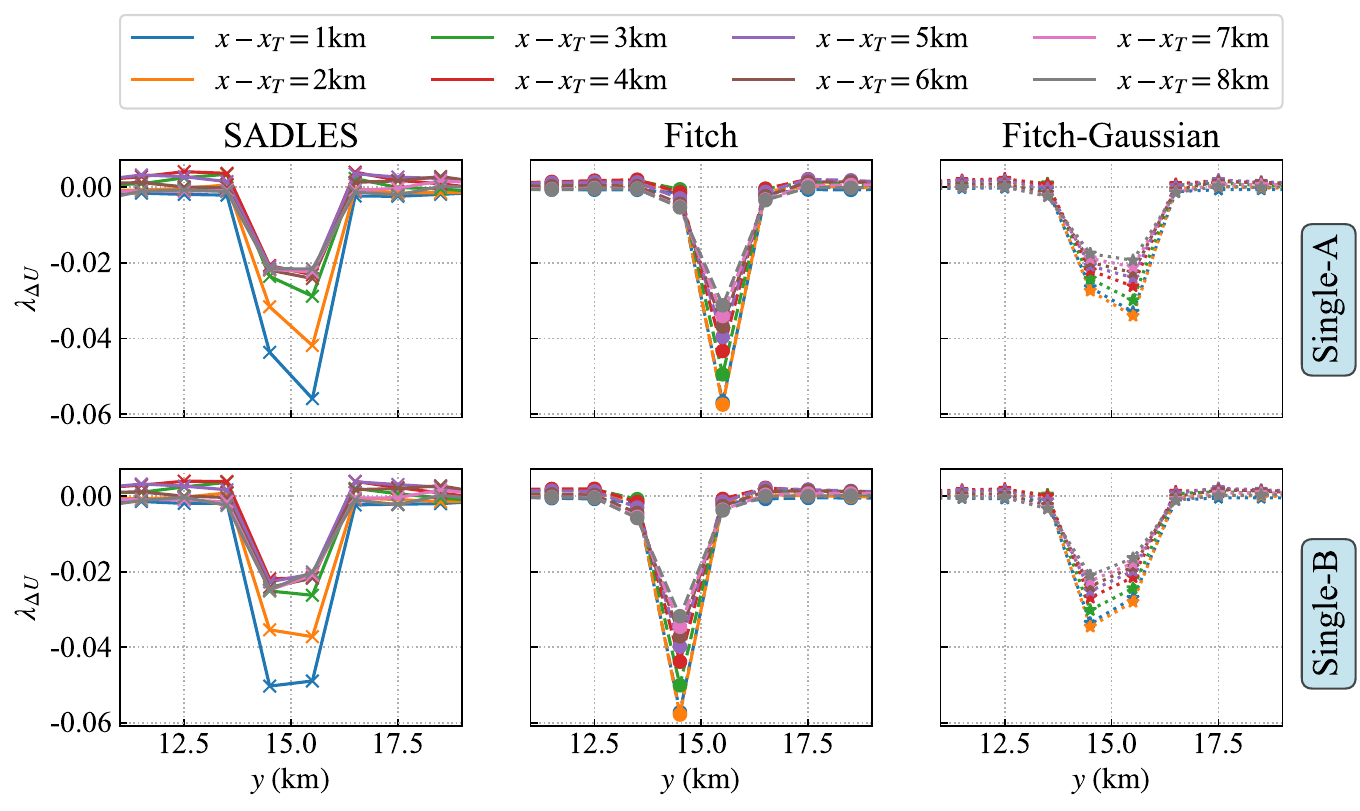}
\caption{\label{fig13} Comparison of the hub-height normalized velocity deficit profiles downstream of the wind turbine for Single-A and Single-B cases obtained from the WRF-LES and WRF-PBL frameworks ($x_T$ is the streamwise location of the wind turbine.)}
\end{figure}
Figure \ref{fig12} shows the hub-height normalized velocity deficits for stand-alone turbine cases (Single-A and Single-B) obtained from WRF-LES and WRF-PBL framework, and the corresponding spanwise profiles at different downstream locations are shown in figure \ref{fig13}. It can be seen that the wake evolution simulated by different WFP models differs significantly due to the different spatial distributions of the momentum sink. The spanwise positions of the wind turbines differ by only 20 m between the Single-A and Single-B cases, but the Fitch model adopts the single-column method, leading to the fact that the sink and source differ by one grid (i.e., 1 km). As a result, the corresponding wind-turbine wakes for the two cases are significantly different, and the peak of the velocity deficit also shifts by one grid, which is far from the results of the WRF-LES framework. Only one velocity deficit peak with a larger magnitude exists and the wakes recover more slowly in the Fitch model. In contrast, the Fitch-Gaussian model can accurately distribute the momentum sink within the two spanwise grids affected by the wind turbine (see figure \ref{fig5}) and capture the dual-peak distribution of the velocity deficit as well as its recovery, demonstrating better prediction accuracy.

\begin{figure}
\centering
\includegraphics[width=0.9\textwidth]{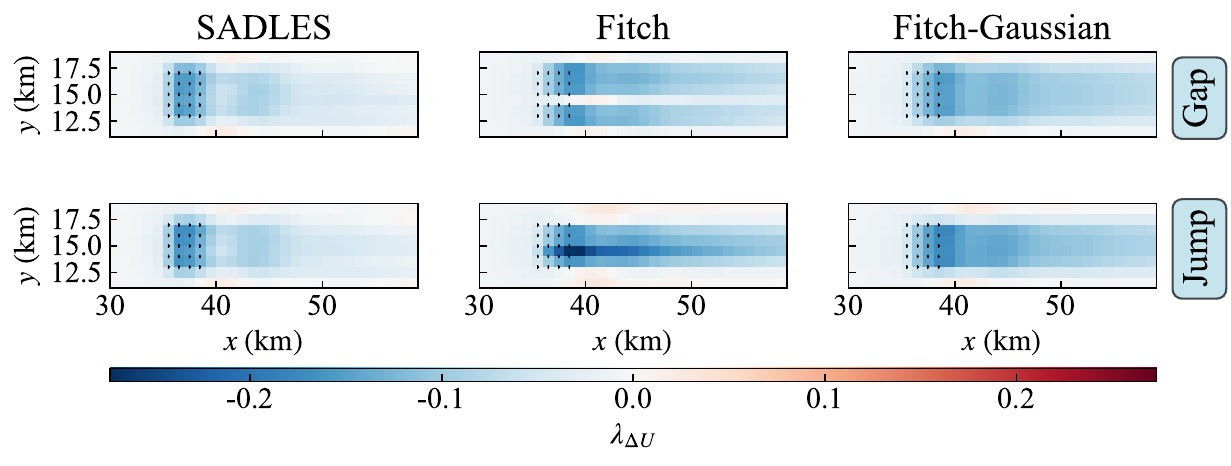}
\caption{\label{fig14} Comparison of the hub-height normalized velocity deficits at hub height for Gap and Jump cases obtained from the WRF-LES and WRF-PBL frameworks}
\end{figure}

\begin{figure}
\centering
\includegraphics[width=0.9\textwidth]{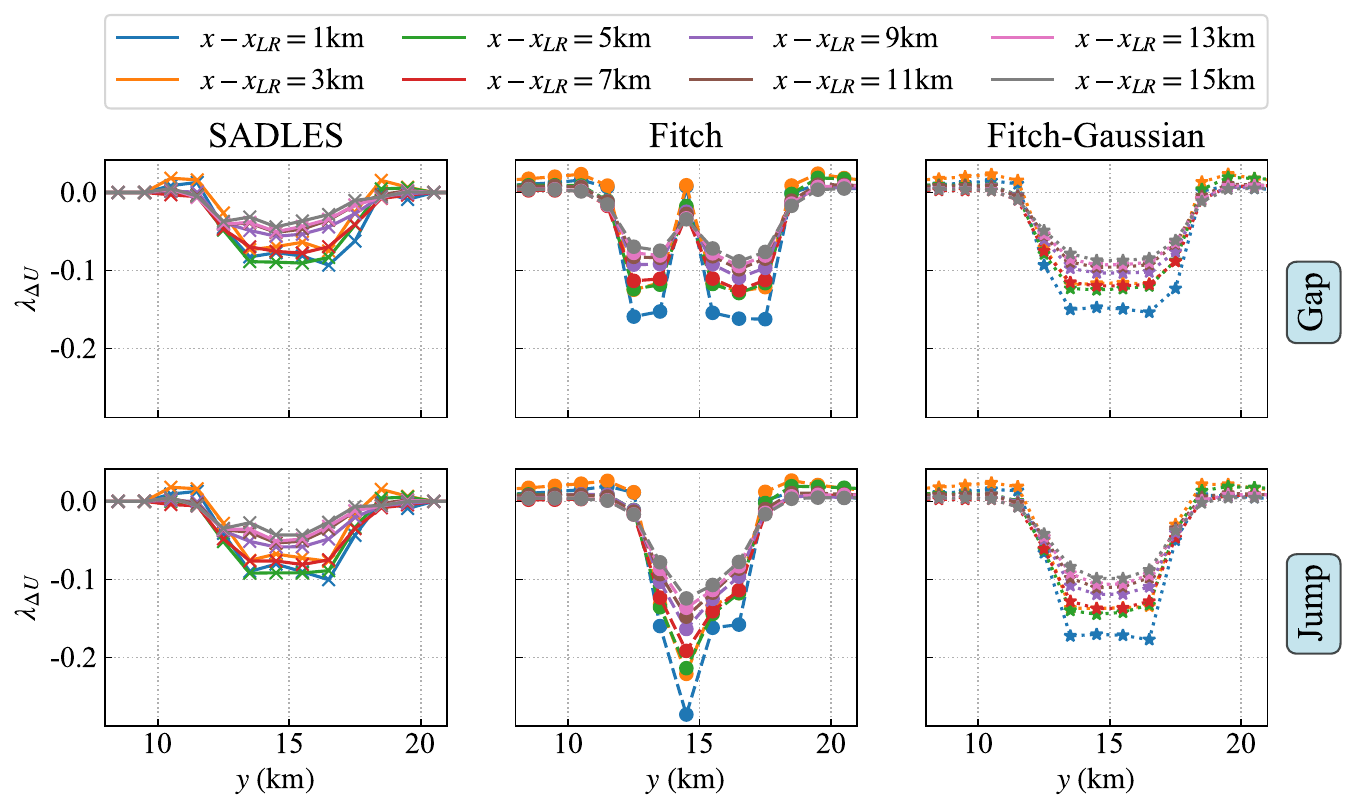}
\caption{\label{fig15} Comparison of the hub-height normalized velocity deficit profiles downstream of the wnid farm for Gap and Jump cases obtained from the WRF-LES and WRF-PBL frameworks ($x_{LR}$ is the streamwise location of the last wind turbine row.)}
\end{figure}

Figure \ref{fig14} shows the hub-height normalized velocity deficits for wind farm cases obtained from the WRF-LES and WRF-PBL frameworks. The corresponding hub-height velocity profiles at different downstream locations of the wind farm are further shown in figure \ref{fig15}. In the Gap (Jump) wind farm case, the Fitch model's momentum sink exhibits an abrupt gap (jump), resulting in high-speed (low-speed) streaks in the wind-farm wakes, which is clearly visible even after 15km. This is significantly different from the mesoscale grid-averaged LES results. In contrast, the Fitch-Gaussian models with the proposed method predict a continuous spanwise distribution of the momentum sink (see figure \ref{fig8}). Therefore, the simulated wind-farm wakes exhibit a nearly top-hat distribution without unphysical streaks. This spatial distribution pattern is much closer to the mesoscale grid-averaged LES results and consistent with the LES simulation results of Lanzilao et al. \cite{46lanzilao2025wind}, demonstrating the advantages of the proposed method in mesoscale simulations. Nevertheless, the wake velocity deficit predicted by the WFP models remains significantly higher than the mesoscale grid-averaged LES results, which is consistent with the finding of Radünz et al. \cite{34radunz2025under}. This is mainly due to the higher momentum sink imposed by the Fitch and Fitch-Gaussian models (see figure \ref{fig8}). More accurate prediction of the momentum sink is expected to significantly lower the simulation error of wind-farm wakes.
\begin{figure}
\centering
\includegraphics[width=0.9\textwidth]{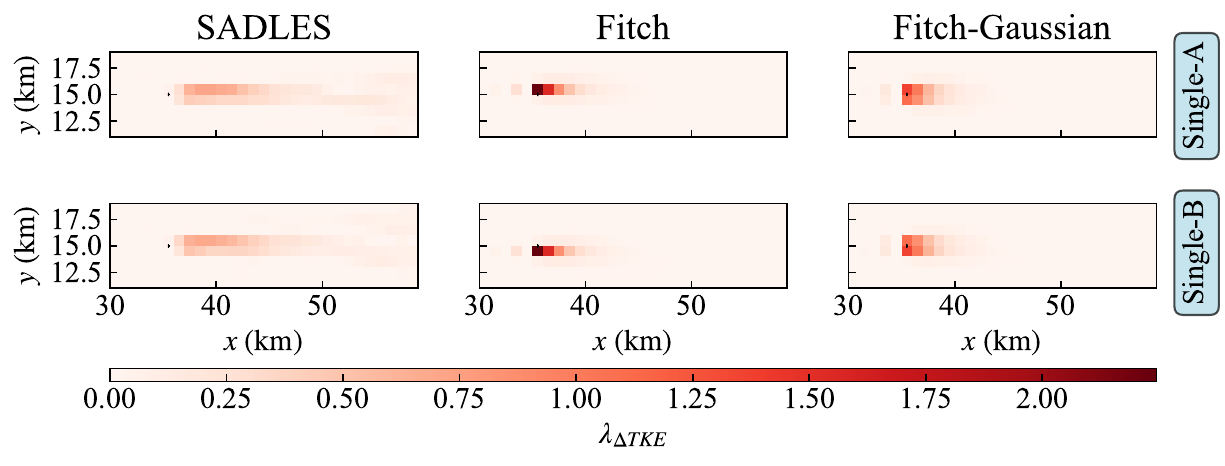}
\caption{\label{fig16} Comparison of the hub-height normalized added TKE for Single-A and Single-B cases obtained from the WRF-LES and WRF-PBL frameworks}
\end{figure}

\begin{figure}
\centering
\includegraphics[width=0.9\textwidth]{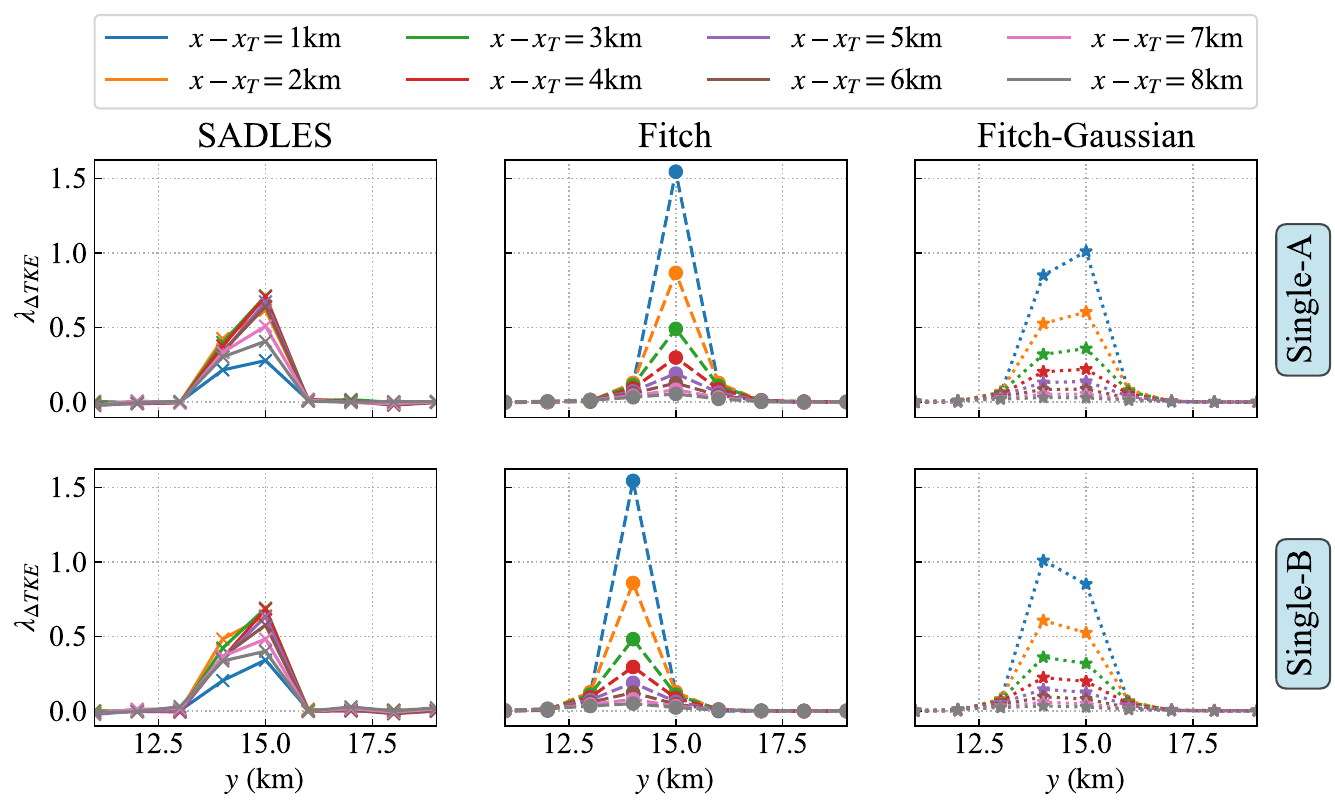}
\caption{\label{fig17} Comparison of the hub-height normalized added TKE profiles downstream of the wind turbine for Single-A and Single-B cases obtained from the WRF-LES and WRF-PBL frameworks ($x_T$ is the streamwise location of the wind turbine.)}
\end{figure}
\subsubsection{\label{sec422} Comparison of added turbulence kinetic energy}
\begin{figure}
\centering
\includegraphics[width=0.9\textwidth]{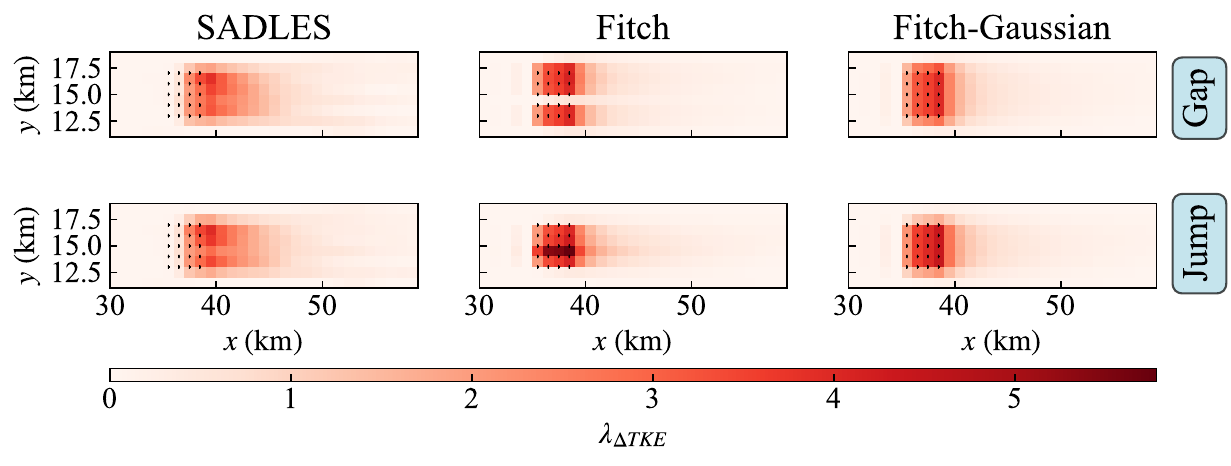}
\caption{\label{fig18} Comparison of the hub-height normalized added TKE for Gap and Jump cases obtained from the WRF-LES and WRF-PBL frameworks}
\end{figure}

\begin{figure}
\centering
\includegraphics[width=0.9\textwidth]{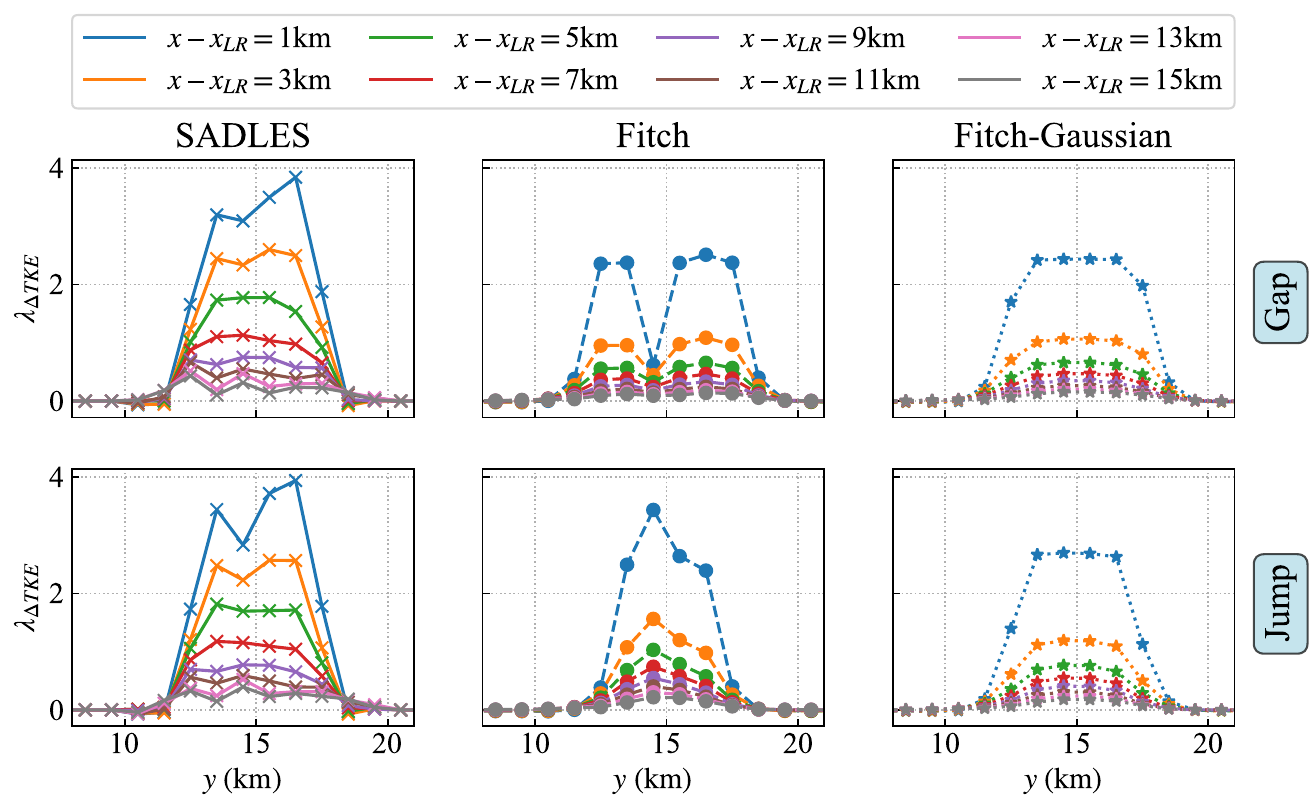}
\caption{\label{fig19} Comparison of the hub-height normalized added TKE profiles downstream of the wind farm for Gap and Jump cases obtained from the WRF-LES and WRF-PBL frameworks ($x_{LR}$ is the streamwise location of the last wind turbine row.)}
\end{figure}
In this section, we further compare the contourfs and spanwise profiles of the added TKE in different cases obtained from the WRF-LES and WRF-PBL frameworks, as shown in figures \ref{fig16} to \ref{fig19}. Similar to velocity deficit, the added TKE in the Fitch model for the stand-alone turbine cases exhibits a single-peak distribution pattern due to the concentrated distribution of the TKE source. In the Gap (Jump) case, the added TKE in the Fitch model exhibits a spanwise gap (jump), which differs significantly from the WRF-LES results. In contrast, the Fitch-Gaussian model using the Gaussian-based multi-column method shows a closer spatial distribution, effectively predicting the impact of wind turbines on the surrounding grids with lower prediction errors.

Although the overall spatial distribution patterns are closer, the magnitude of the added TKE and its streamwise evolution in the mesoscale model still diff significantly from the WRF-LES results. As the TKE source is only concentrated in the grid directly affected by wind turbines, the added TKE is higher in the grid where the wind turbine is located and in the near-wake region. In contrast, the TKE in the far-wake region is only affected by the upstream TKE convection and the vertical wind speed shear production term, and it decays quickly to the ambient TKE. This differs significantly from the physical mechanism of the generation of added TKE in the wind-turbine wakes in the WRF-LES framework. In the WRF-LES framework, wake velocity shear (including spanwise and vertical) is the main production term of TKE. Therefore, the maximum added TKE is not located in the grid where the wind turbine is located, and its peak value generally appears in the wake region. During the downstream evolution of TKE, the velocity shear in the wake region of the WRF-LES framework is significantly higher than the resolved vertical shear in the mesoscale simulation, and the corresponding TKE generation term is also larger, allowing the high TKE region to persist longer in the WRF-LES framework. The fast TKE decay in the mesoscale simulation may affect the wake recovery. As pointed out by Radünz et al. \cite{34radunz2025under} and Bastankhah et al. \cite{47bastankhah2024fast}, the magnitude of the TKE in the wake region directly determines the turbulent transport in the wake region. The higher the TKE, the stronger the turbulent transport, and the faster the wake recovery. Therefore, the fast TKE decay in the mesoscale model may lead to a slow wake recovery \cite{34radunz2025under}.

Based on the above analysis, we can see that although the TKE source in the WFP model is not significantly different from the LES results (see table 4), the added TKE predicted by the mesoscale simulations decays rapidly due to the unresolved velocity shear in the wake region. This suggests that the modeling method of the TKE source in existing WFP models still requires further improvement. To address this problem, Garcia Santiago \cite{39garcia2024mesoscale} introduced the concept of latent kinetic energy, which imposes the TKE source in the region not directly affected by the wind turbines to characterize the contribution of latent kinetic energy. In addition, Archer et al. \cite{48archer2025new} integrated the microscale analytical added TKE model proposed by Khanjari et al. \cite{49khanjari2025analytical} into the Fitch model, which directly gave the magnitude of TKE without calculating the TKE source. These are both very meaningful attempts and may help solve the problem of fast TKE decay and slow velocity recovery in existing WFP models. However, proposing a new method for calculating the TKE source is far beyond the scope of this study. In the future, the above TKE source modelling methods can be combined with the proposed Gaussian-based multi-column method to further improve the simulation accuracy of wind-farm wakes. 
\section{\label{sec5}Conclusion}
Wind farm parameterizations (WFPs) are crucial physical modules for quantifying the impact of wind farms on the atmospheric boundary layer, where wind turbines are typically modeled as the momentum sink and TKE source. Existing WFP models generally use the single-column method to distribute the sink and source to mesoscale grids, which is determines only based on the location of the rotor center. However, this method fails to account for the effects of rotor center variations within the mesoscale grid and neglects the contributions of wind turbines located at grid boundaries to surrounding grids. Therefore, this method tends to lead to concentrated spatial distribution patterns (e.g., gaps or jumps) of the sink and source, thereby affecting the simulation accuracy of wind-farm wakes. To address this, we propose a novel multi-column spatial distribution method based on the Gaussian function for the momentum sink and TKE source in the WFP models. This method uses the rotor center as the origin and distributes the momentum sink and TKE source into multiple vertical grid columns affected by the turbine, with grid weights analytically determined using a mesoscale grid-integrated two-dimensional Gaussian function.

We integrated the proposed method into the Fitch model (based on the mesoscale grid wind speed), proposed the refined Fitch-Gaussian model, and implemented it into the mesoscale WRF model. Based on the high-fidelity WRF-LES cases, we set up stand-alone wind turbine and wind farm cases with wind turbines located at grid boundaries, and comprehensively compared the performance of the proposed method and the conventional single-column method. The results show that, compared with the Fitch model, which relies on the single-column method, the proposed Fitch-Gaussian model can more accurately capture the spatial distribution of the momentum sink and TKE source induced by wind turbines, with lower spatial averaged normalized root mean square error ($NRMSE$) and higher correlation coefficient ($R$). The averaged $NRMSE$ for the sink and source across all the four simulated cases decreases from 0.52 and 0.51 to 0.16 and 0.16. The averaged $R$ for the sink and source across all simulated cases increases from 0.52 and 0.61 to 0.85 and 0.95. More accurate spatial distribution of the sink and source ensure accurate simulation of the velocity deficits and added TKE in wind-turbine/wind-farm wakes. In the stand-alone wind turbine cases, the Fitch model predicts the concentration of the sink and source within a single vertical grid column at the center of the rotor. Consequently, the simulated velocity deficit and added TKE exhibit a single-peak distribution, with a significantly narrow spanwise influence extent. In contrast, the Fitch-Gaussian model captures the spatial distribution of velocity deficit and added TKE in the wind-turbine wakes very well, closely matching the reference mesoscale grid-averaged LES results. In the wind farm cases, the spanwise spatial distribution of sink and source terms predicted by the Fitch model exhibits significant abrupt changes (gaps or jumps), resulting in unphysical streaks in the velocity deficit and added TKE in the simulated wind-farm wakes. In contrast, the Fitch-Gaussian model predicts a roughly top-hat distribution of velocity deficit and added TKE in the wake region, demonstrating a satisfactory agreement with the reference mesoscale grid-averaged LES results.

Although the overall wake spatial distribution characteristics calculated by the Fitch-Gaussian model are similar to LES results, there are still some differences in the magnitude and streamwise evolution of the velocity deficit and added TKE when compared to the LES results. These are inherent limitations of the Fitch model and are beyond the scope of this study. However, this provides two implications for our future research and will serve as interesting directions. First, the velocity deficit in the wind-farm wakes in the mesoscale simulations is higher than the LES results. This is mainly because the momentum sink is calculated directly using the mesoscale grid wind speed, leading to significantly higher results for wind turbines affected by wind-turbine wakes. In the future, the proposed spatial distribution method can be combined with the meso-microscale coupled WFP to address this issue, which uses the rotor-equivalent inflow wind speed to calculate the momentum sink; Second, the TKE in the wake region of the mesoscale simulations decays too quickly. This is primarily due to the unresolved spanwise velocity shear in the mesoscale grid, leading to a low TKE production term in the wake region. A physics-based TKE source modeling method can be developed to account for the unresolved shear, further improving the modeling accuracy of TKE source and slowing TKE decay in the wake region in mesoscale simulations.

\section*{Acknowledgments}
This research was supported by Beijing Natural Science Foundation (No. JQ22008) and the National Natural Science Foundation of China (No. 12172128).

\section*{Author Contributions}
Bowen Du: Conceptualization, Software, Data curation, Writing - original draft.
Mingwei Ge: Conceptualization, Methodology, Supervision, Writing - review \& editing.
Xintao Li: Writing - review \& editing.
Yongqian Liu: Supervision, Writing - review \& editing.
\section*{Data Availability Statement}
The data that support the findings of this study are available from the corresponding author upon reasonable request.

\bibliographystyle{unsrt}
\bibliography{arxiv}  

\end{document}